\documentclass[pdftex, 12pt, a4paper]{article}
\usepackage[pdftex]{graphicx}
\usepackage{epstopdf}
\usepackage[margin=0.8in]{geometry}
\usepackage[english]{babel}
\usepackage[justification=justified,font=small,labelfont=small]{caption}
\usepackage{subcaption}
\usepackage{appendix, parskip, inputenc, tocbibind, url}
\usepackage{wrapfig, enumitem, multirow, tabularx}
\usepackage{amsmath, amssymb, amsthm, braket, empheq, fixltx2e, commath, mathtools}
\usepackage{cleveref}
\usepackage{ragged2e}
\usepackage{gensymb}
\usepackage{float}
\usepackage{cite}
\usepackage[ddmmyyyy]{datetime}

\makeatletter
\newtheoremstyle{indented}
{6pt}
{}
{\addtolength{\@totalleftmargin}{3.5em}
   \addtolength{\linewidth}{-3.5em}
   \parshape 1 3.5em \linewidth}
{}
{\bfseries}
{.}
{.5em}
{}
\newtheoremstyle{indenteditalic}
{6pt}
{}
{\em
   \addtolength{\@totalleftmargin}{3.5em}
   \addtolength{\linewidth}{-3.5em}
   \parshape 1 3.5em \linewidth}
{}
{\bfseries}
{.}
{.5em}
{}
\newtheoremstyle{proof}
{6pt}
{}
{\addtolength{\@totalleftmargin}{3.5em}
   \addtolength{\linewidth}{-3.5em}
   \parshape 1 3.5em \linewidth}
{}
{\em}
{.}
{.5em}
{}
\makeatother

\theoremstyle{indenteditalic}

\numberwithin{thm}{section}

\theoremstyle{indenteditalic}

\theoremstyle{indenteditalic}

\theoremstyle{indented}

\theoremstyle{indented}

\theoremstyle{proof}

\theoremstyle{proof}

\theoremstyle{proof}

\let\bg\begin

\newcommand{\bal}{\bg{align}}

\newcommand{\bsubs}{\bg{subequations}}
\newcommand{\esubs}{\end{subequations}}
\newcommand{\barr}{\begin{array}}
\newcommand{\earr}{\end{array}}

\let\tn\textnormal

\let\tb\textbf
\let\tsups\textsuperscript

\let\mcal\mathcal

\let\lt\left
\let\rt\right
\let\fr\frac

\let\wh\widehat

\let\p\partial

\let\kt\ket
\let\brk\braket

\let\Brk\Braket

\let\a\alpha

\let\Gm\Gamma
\let\gm\gamma
\let\de\delta
\let\De\Delta
\let\O\Omega
\let\o\omega

\let\s\sigma
\let\th\theta
\let\eps\epsilon

\let\ph\phi

\let\Ps\Psi
\let\ps\psi
\let\lm\lambda

\let\t\tau
\let\hb\hbar
\let\rh\rho
\let\k\kappa

\renewcommand{\d}{\tn{d}}

\newcommand{\hH}{\wh{H}}
\newcommand{\hHe}{\wh{H}_\tn{e}}
\newcommand{\hHp}{\wh{H}_\tn{l}}
\newcommand{\hHint}{\wh{H}_\tn{int}}
\newcommand{\ha}{\wh{A}}
\newcommand{\had}{\wh{A}^{\dagger}}

\newcommand{\ktPs}{\kt{\Ps}}

\newcommand{\ktPst}{\kt{\Ps(t)}}

\newcommand{\ktPset}{\kt{\Ps_\tn{e}(t)}}

\newcommand{\ktPspt}{\kt{\Ps_\tn{a}(t)}}

\newcommand{\SE}{Schr\"{o}dinger equation}
\newcommand{\ang}{\r{A}}
\newcommand{\ahelix}{$\a$-helix}
\newcommand{\Romanone}{\uppercase\expandafter{\romannumeral1}}

\newcommand{\Romantwo}{\uppercase\expandafter{\romannumeral2}}

\newcommand{\amideone}{amide-\Romanone}

\newcommand{\brkH}{\brk{H}}
\newcommand{\DT}{\De T}

\newcommand{\zeroN}{\sum_{n=0}^N}
\newcommand{\zeroNmone}{\sum_{n=0}^{N-1}}

\newcommand{\dotpsn}{\dot{\psi}_n}

\newcommand{\tps}{\widetilde{\ps}}
\newcommand{\maxpsnsq}{\max{\abs{\ps_n}^2}}
\newcommand{\BE}{E_\tn{b}}
\newcommand{\Ac}{A_\tn{c}}
\newcommand{\hQ}{\widehat{Q}}
\newcommand{\hP}{\widehat{P}}

\newcommand{\hPd}{\widehat{P}^{\dagger}}
\newcommand{\hs}{\widehat{\s}}
\newcommand{\hsd}{\hs^{\dagger}}
\newcommand{\ktvace}{\kt{0_{\tn{e}}}}
\newcommand{\ktvacp}{\kt{0_{\tn{a}}}}
\newcommand{\sech}{\tn{sech}}


\bg{document}
\pagestyle{plain}
\setlength{\baselineskip}{14.4pt}


\begin{titlepage}
\begin{center}

\vspace*{60pt}
\LARGE{\tb{Directed polaron propagation in linear polypeptides induced by intramolecular vibrations and external electric pulses}}

\vspace{14.4pt}
\large{DOI: 10.1103/PhysRevE.98.012401}

\vspace{60pt}
\large{\textsc{J. Luo}*}

\large{\textsc{B. M. A. G. Piette}$\dagger$}

\vfill
\large{*jingxi.luo@durham.ac.uk}

\large{$\dagger$b.m.a.g.piette@durham.ac.uk}

\vspace{14.4pt}
\large{Department of Mathematical Sciences, Durham University \\ Durham, DH1 3LE, United Kingdom}

\end{center}
\end{titlepage}


\begin{titlepage}
\begin{center}

\begin{minipage}{\textwidth}

\vspace*{120pt}
\begin{flushleft}
\Large{\textsc{Abstract}}
\end{flushleft}

\vspace{14.4pt}
We study the propagation of $\alpha$-helix polarons in a model describing the non-adiabatic interaction between an electron and a lattice of quantum mechanical oscillators at physiological temperature. We show that when excited by a sub-picosecond electric pulse, 
as induced by experimentally observed sub-picosecond charge separation, the polaron is displaced by up to hundreds of lattice sites before the electron becomes delocalised. We discuss biophysical implications of our results.

\end{minipage}
\end{center}
\end{titlepage}


	\section{Introduction} \label{intro}

Directed motion of electrons in proteins is important to a wide range of biological processes, from cellular respiration to photosynthesis \cite{Feher1989,Senior2002,Karp2008,Kaucikas2016}. Past studies have tended to examine the transport process in the framework of quantum tunnelling \cite{Beratan1987,Beratan1991,Beratan1992,Hasegawa2017}. More recently, reports have emerged in which electron transport in either linear or helical polypeptides is modelled as a polaron effect \cite{Hennig2001,Brizhik2008,Brizhik2010,Brizhik2014,Luo2017}. That is, the electron potential is modulated by the vibrational dynamics of the protein molecule, and this interaction induces a deep potential well, trapping the electron. The ``self-trapped'' electron and the local molecular distortion form a quasi-particle compound known as a polaron, which is a familiar concept in solid state physics \cite{Landau1933,Pekar1946,Frohlich1952,Holstein1959}. If the electron is to propagate in a lossless solitonic fashion, as opposed to dispersing and becoming a free particle, it must move in sync with the local molecular distortion. Thus, the integrity of the polaron remains intact.

It has been shown that, once a static polaron is formed on a one-dimensional molecular lattice, transport is possible after a kick in the lattice pinning mode \cite{Hennig2001}, or under the influence of a biharmonic electric field with zero mean \cite{Brizhik2008,Brizhik2010,Brizhik2014} or harmonic electric field with non-zero mean \cite{Luo2017}. In the current study, we consider polarons formed in $\alpha$-helicies by electrons interacting with intramolecular vibrations. We consider both thermalised and non-thermalised lattices. Our aim is to find a suitable external forcing which facilitates directed electron motion along the lattice, and to describe the characteristics of such motion. We discover that an electric pulse with appropriate amplitude and hundred-femtosecond timespan can be used to displace a polaron by tens of lattice sites. Such electric pulses
match exactly, both in amplitude and timespan, those induced by the charge 
separation observed in biological complexes reported in 
\cite{Gauduel1987,Zhong2001}.
When these pulses are repeated periodically in time, we find that the polaron can remain intact for several periods, during which it can be displaced by hundreds of sites. A similar excitation was described in \cite{Lakhno2011} for DNA polarons but without thermal effects. 

In \Cref{section2}, we outline our mathematical model for the non-adiabatic interaction between an electron and lattice vibrations, and describe the physical parameters, and we derive a set of dynamical equations from the model. 
In \Cref{section3} we present solutions representing stationary polaron states, comparing numerical and analytical results. \Cref{section4} concerns propagating solutions, where we find suitable electric fields capable of displacing a polaron from its stationary state and sustaining its motion. We examine the effect of a single electric pulse as well as periodically repeated pulses, and we characterise the resulting polaron motion in terms of velocity and stability. By taking thermal effects into account, we investigate the stability of the polaron with respect to random forces due to temperature in the environment. At physiological temperature, not only is the polaron dynamics stable, but also the random forces promote directed transport in the sense that, when thermal effects are present, the polaron has a stronger tendency to move in one direction over the other. 

Throughout this study, values of physical parameters are chosen to correspond to an electron interacting with intramolecular oscillators in a hydrogen-bonded peptide chain in an \ahelix. The most common secondary protein structure, the \ahelix~consists of amino acid residues linked by hydrogen bonds \cite{Pauling1951,Dunitz2001}, and it has various intermolecular and intramolecular vibrational degrees of freedom, most notably the stretching of the hydrogen bonds \cite{Brown1972}, and various intramolecular vibrational modes such as \amideone~vibrations in C=O double bonds
\cite{Nevskaya1976b}. We have chosen to consider the C=O double oscillator as the unit of our lattice, and this determines the key parameter values in our model.


	\section{The model and dynamical equations} \label{section2}

We consider a linear molecular lattice with unit mass $M$ and equilibrium spacing $R$. In the absence of extraneous electrons, every unit cell independently oscillates with natural angular frequency $\O$. We model the lattice by quantum mechanical operators as opposed to classical variables, because we wish to assign a specific frequecy $\O$ corresponding to the \amideone~mode, and it has been shown that the absorption band of a theoretical classical \amideone~oscillator is 40 times wider than the quantum mechanical one \cite{Cruzeiro-Hansson1997}. We consider \amideone~vibration excitations on the lattice points, which could be induced by a passing electron for example, and describe the system using a Fr\"{o}hlich-Holstein Hamiltonian, $\hH = \hHe + \hHp + \hHint$, where
\bsubs \label{DSham}
\bal
\hHe &=  \sum_{n=0}^N J_0 \had_n \ha_n - \sum_{n=0}^{N-1} J_1 \lt(\had_{n+1}\ha_n + \had_n\ha_{n+1}\rt) - \zeroN e E(t) R \lt( n - n_0 \rt) \had_n \ha_n, \label{DSham_e} \\
\hHp &= \sum_{n=0}^N \lt( \fr{{\hP_n}^2}{2M} + \fr{M \O^2 \hQ_n^2}{2} \rt), \label{DSham_p} \\
\hHint &= \sum_{n=0}^{N} \chi \hQ_n \had_n\ha_n, \label{DSham_int}
\end{align}
\esubs
In \cref{DSham}, $n = 0, \dots, N$ labels the lattice sites. $\hHe$ describes a tight-binding electron, where $\had_n$ and $\ha_n$ are, respectively, the operators of electron creation and annihilation at the $n\tsups{th}$ lattice site. $J_0$ is the potential energy of a stationary electron in a transfer-free and distortion-free lattice, and $-J_1$ is the electron exchange energy. $-e$ is the electron charge, $E(t)$ the amplitude of an external electric field, and $n_0$ an arbitrary lattice site at which the potential energy due to $E(t)$ is set to zero. The last term in \cref{DSham_e} represents the modification of on-site electron energies due to the presence of the electric field \cite{Brizhik2014}. $\hHp$ corresponds to the energy contribution from the lattice, where $\hQ_n$ and $\hP_n$ are, respectively, the displacement and conjugate momentum operators for the $n\tsups{th}$ oscillator. The form of the interaction Hamiltonian, $\hHint$, is derived from the assumption that on-site electron energies are modulated by the displacement field, $\hQ_n$, and we have retained only the linear term, involving coupling constant $\chi$, in the Taylor expansion for this modified energy \cite{Hennig2001}. The operators satisfy the commutation relations
\bsubs \label{commutation}
\bal
&\lt[ \hQ_m, \hP_n \rt] = - \lt[ \hQ_m, \hPd_n \rt] = i \hb ~\delta_{mn}, \label{commute1} \\
&\lt[ \hQ_m, \ha_n \rt] = \lt[ \hQ_m, \had_n \rt] = 0 = \lt[ \hP_m, \ha_n \rt] =\lt[ \hP_m, \had_n \rt], \label{commute2}
\end{align}
\esubs
and the fermionic anti-commutation relation
\bal
\ha_m \had_n + \had_n \ha_m = \delta_{mn}. \label{anticommute}
\end{align} 
Denoting the vacuum states of $\hHe$ and $\hHp$ by, respectively, $\ktvace$ and $\ktvacp$, we have $\ha_n \ktvace = 0$ and $\hQ_n \ktvacp = \hP_n \ktvacp = 0$. Then at time $t$ the electronic state is a superposition of single excitations, $\ktPset = \sum_{n=0}^N \a_n(t) \had_n \ktvace$ for some complex coefficients $\a_n$. We are assuming that the intramolecular oscillators are in a Glauber state \cite{Eilbeck1984,Kerr1987}, $\ktPspt = \exp(\hs(t)) \ktvacp$, where
\bal
\hs(t) = \fr{i}{\hb} \sum_{n=0}^N \lt( p_n(t) \hQ_n - q_n(t) \hP_n \rt), \label{defn_hs}
\end{align}
for some real coefficients $p_n, q_n$. The Glauber state is a pure state and therefore does not account for entanglement effects, so our model is semi-classical despite the appearance of the $\hP_n, \hQ_n$ operators. We further assume that the electron and amide subsystems are not entangled, so that the state of the composite system can be written
\bal
\ktPst = \sum_{n=0}^N \a_n(t) \exp ( \hs(t) )~ \had_n \ktvace \ktvacp, \label{defn_ktPst}
\end{align}
with the normalisation condition,
\bal
\zeroN \abs{\alpha_n}^2 = 1. \label{norm_cond}
\end{align}
Using \cref{commutation}, and the fact that $\hsd = - \hs$, as well as the Baker-Hausdorff identity for quantum operators, we derive
\bsubs
\bal
\exp(\hsd) \hQ_n \exp(\hs) &= \hQ_n + q_n, \qquad \exp(\hsd) \hQ_n^2 \exp(\hs) = \hQ_n^2 + 2q_n \hQ_n + q_n^2, \\
\exp(\hsd) \hP_n \exp(\hs) &= \hP_n + p_n, \qquad \exp(\hsd) \hP_n^2 \exp(\hs) = \hP_n^2 + 2p_n \hP_n + p_n^2.
\end{align}
\esubs
It therefore follows that the expected value of the total energy of the system in state $\ktPs$, $\brk{H} := \brk{\Ps | \hH | \Ps}$, is
\bsubs
\bal
\brk{H} &= \sum_{j=0}^N \sum_{k=0}^N \a_j^* \a_k \Brk{ 0 | \ha_j \exp(\hsd) \hH \exp(\hs) \had_k | 0 } \label{H1} \\
&= \sum_{j=0}^N \sum_{k=0}^N \a_j^* \a_k \Brk{ 0 | \ha_j \lt( \hHe + \sum_{m=0}^N \Big( \fr{p_m^2}{2M} + \fr{M\O^2  q_m^2}{2} + \chi q_m \had_m \ha_m \Big) \rt) \had_k | 0 }, \label{H2}
\end{align}
\esubs
where $\kt{0} = \ktvace \ktvacp$. To derive dynamical equations for $\a_n$ and $q_n$, we proceed as follows. For $q_n$, we have $\d q_n / \d t = \p \brkH / \p p_n$ and $\d p_n / \d t = - \p \brkH / \p q_n$, the combination of which gives
\bal
M \fr{\d^2 q_n}{\d t^2} = - \lt( M \O^2 q_n + \chi \abs{\a_n}^2 \rt). \label{eqn_ddotqj}
\end{align}
For $\a_n$, we deduce from \cref{H1}, making use of \cref{defn_ktPst} and the \SE~$\hH \ktPs = i\hb~ \p \ktPs / \p t$, that
\bal
\fr{\p \brkH}{\p \a_n^*} = i \hb \sum_{k=0}^N \Brk{0 | \ha_n \exp(\hsd) \lt( \fr{\d \a_k}{\d t} + \a_k \fr{\d \hs}{\d t} \rt) \exp(\hs) \had_k | 0},
\end{align}
At this point, a standard treatment is to invoke the adiabatic approximation, that $\d \hs / \d t$ is negligible compared to $\d \a / \d t$ \cite{Scott1992}. Here we do not make such an assumption, therefore 
\bal
\fr{\p \brkH}{\p \a_n^*} &= i \hb \sum_{k=0}^N   \fr{\d \a_k}{\d t}~\de_{nk} - \sum_{k=0}^N \a_k \Brk{0 | \ha_n \sum_{m=0}^N \lt( \fr{\d p_m}{\d t} \lt( \hQ_m + q_m \rt) - \fr{\d q_m}{\d t} \lt( \hP_m + p_m \rt) \rt) \had_k | 0} \nonumber \\
&= i \hb \fr{\d \a_n}{\d t} + \lt( 2W + I \rt) \a_n,  \label{eqn_an}
\end{align}
where 
\bal
W(t) = \brk{\hHp} = \fr{1}{2} \sum_{m=0}^N \lt( M \Big( \fr{\d q_m}{\d t} \Big)^2 + M \O^2 q_m^2 \rt), \quad I(t) = \brk{\hHint} = \sum_{m=0}^N \chi q_m \abs{\a_m}^2
\end{align}
are the expected energy contributions from the lattice and electron-lattice interaction, respectively. Meanwhile, from \cref{H2},
\bal
\fr{\p \brkH}{\p \a_n^*} &= \sum_{k=0}^N \a_k \Bigg\langle 0~ \Bigg\vert \ha_n \Bigg( \sum_{m=0}^N J_0 \had_m \ha_m - \sum_{m=0}^{N-1} J_1 \lt(\had_{m+1}\ha_m + \had_m\ha_{m+1}\rt)  \nonumber \\
& \qquad \qquad - \sum_{m=0}^N e E(t) R \lt( m - n_0 \rt) \had_m \ha_m +  \sum_{m=0}^N \Big( \fr{p_m^2}{2M} + \fr{M\O^2  q_m^2}{2} + \chi q_m \had_m \ha_m \Big) \Bigg) \had_k \Bigg\vert ~0 \Bigg\rangle \nonumber \\
&= \a_n \lt[ J_0 + \chi q_n - e E R \lt( n - n_0 \rt) + W \rt]  - J_1 \lt( \a_{n-1} + \a_{n+1} \rt). \label{eqn19}
\end{align}
Comparing \cref{eqn_an,eqn19}, we obtain the following equation for $\a_n$.
\bal
 i \hb \fr{\d \a_n}{\d t} = \a_n \lt[ J_0 + \chi q_n - e E R \lt( n - n_0 \rt) - \lt( W + I \rt) \rt]  - J_1 \lt( \a_{n-1} + \a_{n+1} \rt). \label{dyn_eqn_an}
\end{align}
\Cref{eqn_ddotqj,dyn_eqn_an} are the coupled dynamical equations for our system. We have defined $\a_{-1} = \a_{N+1} = 0$, so that \cref{dyn_eqn_an} holds at the boundaries. We note that, had we used the adiabatic approximation, the $-(W+I)$ in \cref{dyn_eqn_an} would have been replaced by $+W$. Next, we define a variable $\ps_n(t)$ with $\abs{\ps_n} = \abs{\a_n}$ by
	\bal
	\a_n (t) = \ps_n (t) \exp \lt[ \fr{it}{\hb} \lt( -J_0 + 2J_1 \rt) \rt], \label{gauge}
	\end{align}
which redefines the zero of energy measurements so that $J_0 = 0$ in \cref{dyn_eqn_an}, and the $J_1$ term becomes a discrete Laplacian, $-J_1 \lt( \a_{n-1} + \a_{n+1} - 2 \a_n\rt)$. To account for the effect on the lattice of its thermal environment, we add two Langevin terms, $-\Gm ~\d q_n / \d t + F_n(t)$, to the r.h.s. of \cref{eqn_ddotqj} \cite{Lemons1997,Schlick2010}. Here, $\Gm$ is a viscous damping coefficient, which depends on temperature of the environment, and $F_n(t)$ are Gaussian stochastic terms describing thermal fluctuations, with zero mean and satisfying the correlation relation, $\brk{F_m (t), F_n (t')} = 2 \Gm k_B T \de_{m,n} \de( t - t' )$, where $k_B$ is the Boltzmann constant and $T$ is the temperature. Scaling time by $\O^{-1}$ and length by $\sqrt{\hb M^{-1} \O^{-1}}$, we have the following non-dimensionalised dynamical equations for $\ps_n$ and dimensionless lattice displacements $u_n$.
	\bsubs \label{dimlesseqns}
	\bal
	i \dotpsn &= \lt( \k u_n - w(\t) - \eta(\t) \rt)\ps_n - \rh \lt( \ps_{n-1} + \ps_{n+1} - 2\ps_n \rt) - \eps(\t) (n - n_0) \ps_n, \label{dimlessphn} \\
	\ddot{u}_n &= - u_n - \k \abs{\ps_n}^2 - \gm \dot{u}_n + f_n(\t), \label{dimlessun}
	\end{align}
	\esubs
where $\t = \O t, u_n = q_n / \sqrt{\hb M^{-1} \O^{-1}}, w = W / (\hb \O), \eta = I / (\hb \O)$, and 
	\bal
	\k = \fr{\chi}{\sqrt{\hb M \O^3}}, \quad \rh = \fr{J_1}{\hb \O}, \quad \eps = \fr{e E R}{\hb \O}, \quad \gm = \fr{\Gm}{M \O}, \quad f_n = \fr{F_n}{\sqrt{\hb M \O^3}}. \label{dimlessparams}
	\end{align}
At the boundary we have $\ps_{-1} = \ps_{N+1} = 0$. Physically, $\rh$ is the characteristic timescale separation between the electron and lattice, which is why it is known in the literature as the \emph{adiabaticity parameter} \cite{Hennig2001}, and $\k$ is a measure of the \emph{coupling strength} between electron and lattice. We take the following parameter values appropriate for a single hydrogen-bonded peptide chain in an \ahelix~\cite{Pauling1951,Dunitz2001,Chothia1977,Barlow1988,Nevskaya1976b,Zavitsas1987,Hennig2001}. In cases where no exact value is known, a suitable range is given. $M = 1.147 \times 10^{-26}$kg, $R = 4.5$\ang, $\O = 3.1 \times 10^{14} \tn{s}^{-1}$, $J_1 \lessapprox 1\tn{eV}$. We treat $\chi$ and $E(t)$ as adjustable parameters in the model, so that $\kappa$ and $\eps(\t)$ are adjustable. Appropriate estimates of $\Gamma$ can be obtained by approximating the lattice units as spheres, for lack of a better method, and using Stokes' Law \cite{Laidler1982}. Its value decreases as temperature is raised, but for the remainder of this study we fix the dimensionless parameter $\gm = 0.001$, which corresponds to $\Gamma$ at physiological temperature, i.e. $T = 310$K. We also have $\rh \lessapprox 5$, and the thermal energy $\th = k_B T / (\hb \O) = 0.13$ enters the system via the stochastic forcing, $f_n(\t)$, which satisfies  
	\bal
	\brk{f_n(\t)} = 0, \qquad \brk{f_m (\t), f_n (\t + \De \t)} = 2 \gm \th \de_{m,n} / \De \t . \label{thermal_term}
	\end{align}
We investigate the effect of $f_n$ in \Cref{section3,section4} by comparing results of solving the system neglecting $f_n$ and solving the system with $f_n$ taken into account.


	\section{Stationary polarons} \label{section3}

With $f_n(\t) = 0$, the stationary solution to \cref{dimlessun} is
	\bal
	u_{n} = - \k \abs{\ps_n}^2. \label{steady_un}
	\end{align}
This reflects the fact that lattice excitations and electronic excitations in our model are directly related, one arising from the presence of the other.  Using a gauge transformation $\ps_n \mapsto \ps_n \exp [i (w+\eta) \t ]$, where $w$ and $\eta$ are constants in the steady state, we set $w+\eta = 0$ in \cref{dimlessphn}. Putting \cref{steady_un}, as well as $\eps(\t) = 0$, into \cref{dimlessphn}, we obtain
	\bal
	i \rho^{-1} \dotpsn + \lt( \ps_{n-1} + \ps_{n+1} - 2\ps_n \rt) + \lm \abs{\ps_n}^2 \ps_n = 0, \label{nls_discrete}
	\end{align}
where 
	\bal
	\lm := \fr{\k^2}{\rh} = \fr{\chi^2}{M \O^2 J_1} \label{defn_lm}
	\end{align}
is known as the \emph{effective coupling parameter} \cite{Alexandrov1995}. Since $\ps_n$ is stationary, the time-evolution of $\ps_n$ manifests only as a variation in its phase. That is, for some eigenvalue $H_0$, we have
	\bal
	\ps_n(\t) = \ph_n \exp \lt( -i \rh  H_0 \t \rt), \label{steady_psn}
	\end{align}
where $\ph_n$ are time-independent. Now $\rh$ appears only as a phase factor, it is immediately clear that the stationary $\abs{\ps_n}^2$ solutions to our system are fully characterised by $\lm$ \cite{Cruzeiro-Hansson2000}.


	\subsection{Continuum limit} \label{subsection_continuum}

 In the limit $N \gg 1$, \cref{nls_discrete} is the spatially-discretised nonlinear \SE~(NLSE). That is, from the NLSE $i\hb~ \p_t \tps + R^2 J_1 \p_x^2 \tps + \lm J_1 |\tps|^2 \tps = 0$ on a domain of size $RN$ with $N \gg 1$, one can obtain \cref{nls_discrete} by writing $\p_x^2 \tps = [\tps(x-R) + \tps(x+R) - 2\tps(x) ]/ R^2$ and identifying $\tps(nR)$ with $\ps_n$. One can therefore obtain an approximate solution to \cref{nls_discrete} by discretising the stationary solution to the NLSE, namely $\tps(x) = \sqrt{\lm/8}~ \sech \lt[ \lm \lt(x - x_0 \rt) / (4 R) \rt] \exp (-i J_1 H_0 t / \hb)$ with eigenvalue $H_0 = -\lm^2 / 16$ \cite{Zakharov1972,Hirota1973,Li2007}. Indeed, the approximate stationary solution obtained via the continuum limit is  $\abs{\ps_n}^2 = \abs{\ph_n}^2 = (\lm/8) ~\sech^2 ~[ \lm ( n - n_0 ) / 4]$.  The half-width of this profile is inversely proportional to $\lm$, whilst the profile height is proportional to $\lm$. 

We note that, as well as stationary solutions, the NLSE also admits mobile solutions in the form of travelling waves. But they are of no interest to us because \cref{nls_discrete} is derived from the system only if the $u_n$ field - an integral part of the polaron - is stationary.


	\subsection{Numerical solutions with $f_n = 0$} \label{subsection_numerics}

Without invoking the continuum limit, we must compute the stationary solutions numerically. To achieve this, we use a numerical shooting method with the $\ph_n \sim \sech(n)$ approximations as initial guess. This gives us $\ph_n$ solutions, from which we can derive all the important information about the stationary polaron state, namely the electron probability density $|\ps_n|^2 = |\ph_n|^2$, the stationary lattice configuration $u_n = -\k | \ps_n |^2$, as well as the \emph{binding energy} of the polaron, $\BE$, which for a general state $\Ps(t)$ is given w.r.t. $J_0$ and in units of $\hb \O$ by
	\bal
	\BE = \fr{\brk{\Ps | \hH | \Ps} - J_0}{\hb \O} = - \zeroNmone \rh \lt( \ps^*_{n+1} \ps_n + \ps^*_n \ps_{n+1} \rt) + \zeroN \lt( \fr{u_n^2}{2} + \fr{\dot{u}_n^2}{2} \rt) + \zeroN \k u_n \abs{\ps_n}^2 ,
	\end{align}
and in the stationary state the binding energy is
	\bal
	\BE^0 = - \zeroNmone \rh \lt( \ps^*_{n+1} \ps_n + \ps^*_n \ps_{n+1} \rt) - \zeroN \fr{\k^2}{2} \abs{\ps_n}^4. \label{exactH0}
	\end{align}  
We note that the gauge transformation given by \cref{gauge} effectively shifts the energy spectrum of the system by the constant $2 J_1 - J_0$, which is why we measure energy from $J_0$ instead of, as is commonly done, the lowest energy in the electron band, $- 2 J_1$.

	\bg{figure}[h]
	\centering
	\bg{subfigure}[h]{0.5\textwidth}
	\includegraphics[width=\textwidth]{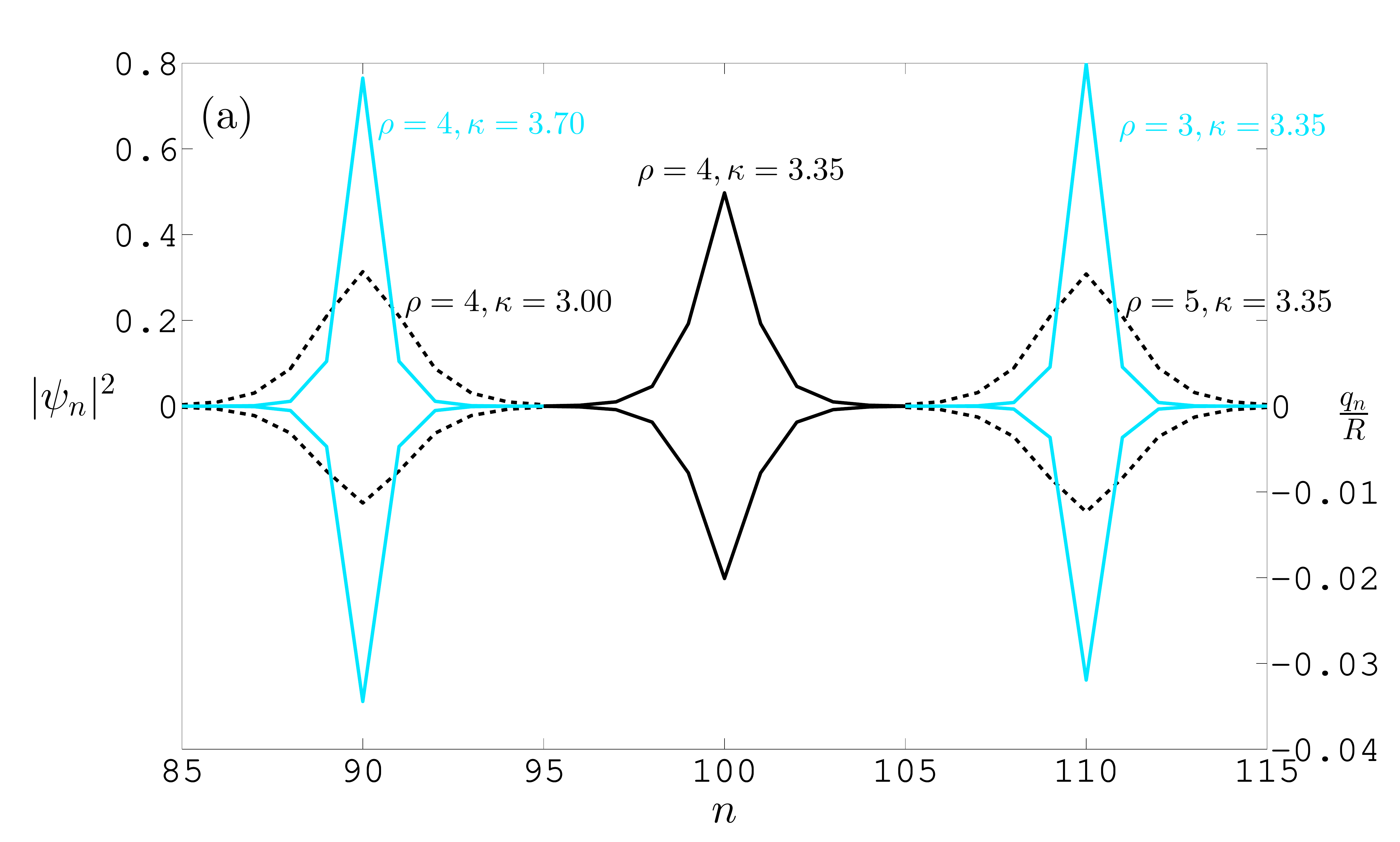}
	\end{subfigure}%
	\bg{subfigure}[h]{0.5\textwidth}
	\includegraphics[width=\textwidth]{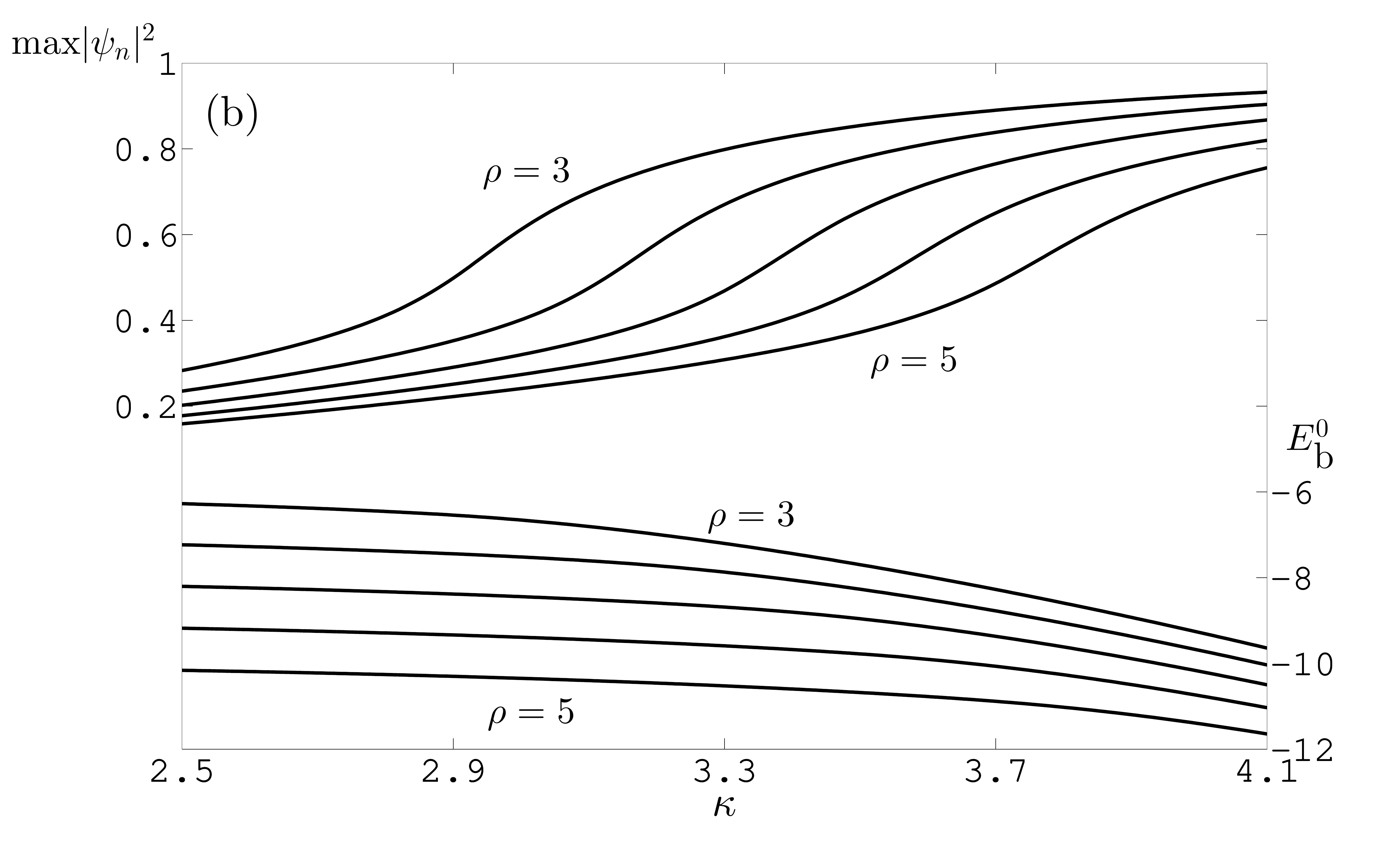} 
	\end{subfigure}%
	\caption{Stationary solutions to \cref{dimlesseqns} with $\eps(\t) = f_n(\t) = 0$. (a) Some $|\ps_n|^2$ solutions (left axis), and corresponding $q_n / R$ which is lattice distortion in units of equilibrium spacing (right axis), computed using various combinations of parameters $\rho$ and $\k$. (b) Dependence on $\rho$ and $\k$ of two key characteristics of stationary polaron states: the maximum localisation probability, $\maxpsnsq$ (left axis), and the binding energy, $\BE^0$ (right axis). Both are expressed as families of functions of $\k$, parametrised by $\rh = 3.0,3.5,4.0,4.5,5.0$.} \label{fig_stat_pol}
	\end{figure}
We have obtained stationary solutions to \cref{dimlesseqns} with $\eps = f_n = 0$ on a grid of size $N=200$. In Figure \ref{fig_stat_pol}(a) we see some $\abs{\ps_n}^2$ profiles, the height of which increases with $\k$ and decreases with $\rh$. This is as we expected, since the stationary solutions are fully characterised by $\lm = \k^2 / \rh$. It also shows $q_n$ being proportional to $| \ps_n |^2$, as it should be. Figure \ref{fig_stat_pol}(b) serves to explain the dependency of the polaron state on $\rh$ and $\k$. For fixed $\rh$, $\maxpsnsq$ increases with $\k$ whilst $\BE^0$ decreases with $\k$. The negative sign of $\BE^0$ signifies that energy needs to be put into the system in order to break up the polaron; thus, a decrease in $\BE^0$ is an indication that the electron is more strongly bound to the lattice. Since a larger $\k$ indicates a stronger electron-lattice interaction, we do expect that it results in a more strongly bound polaron. Meanwhile, for fixed $\k$, $\maxpsnsq$ decreases with $\rh$ and $\BE^0$ decreases with $\rh$. An extension of Figure \ref{fig_stat_pol}(b) is Figure \ref{fig_param_space}(a), where we see $\maxpsnsq$ and $\BE^0$ as surfaces over the parameter space $(\rh,\k)$. By drawing contour lines of the $\maxpsnsq$ surface, we obtain Figure \ref{fig_param_space}(b). Indeed, these contour lines are the parabolae $\k^2 / \rh = $ constant, i.e. lines of constant $\lm$. This means that the shape of a $\abs{\ps_n}^2$ profile depends solely on $\lm$, again as we would expect.
	\bg{figure}[h]
	\centering
	\bg{subfigure}[h]{0.5\textwidth}
	\includegraphics[width=\textwidth]{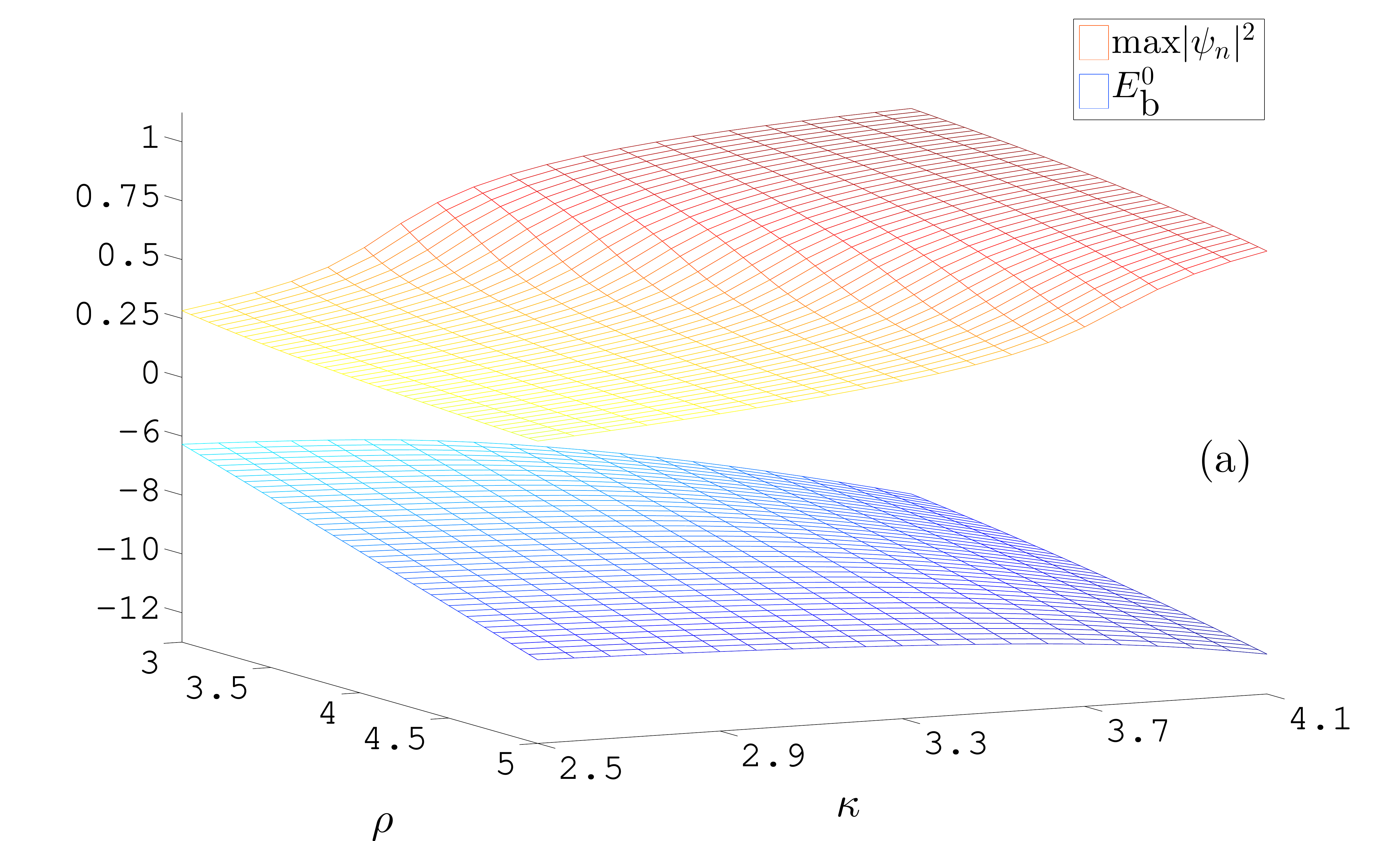}
	\end{subfigure}%
	\bg{subfigure}[h]{0.5\textwidth}
	\includegraphics[width=\textwidth]{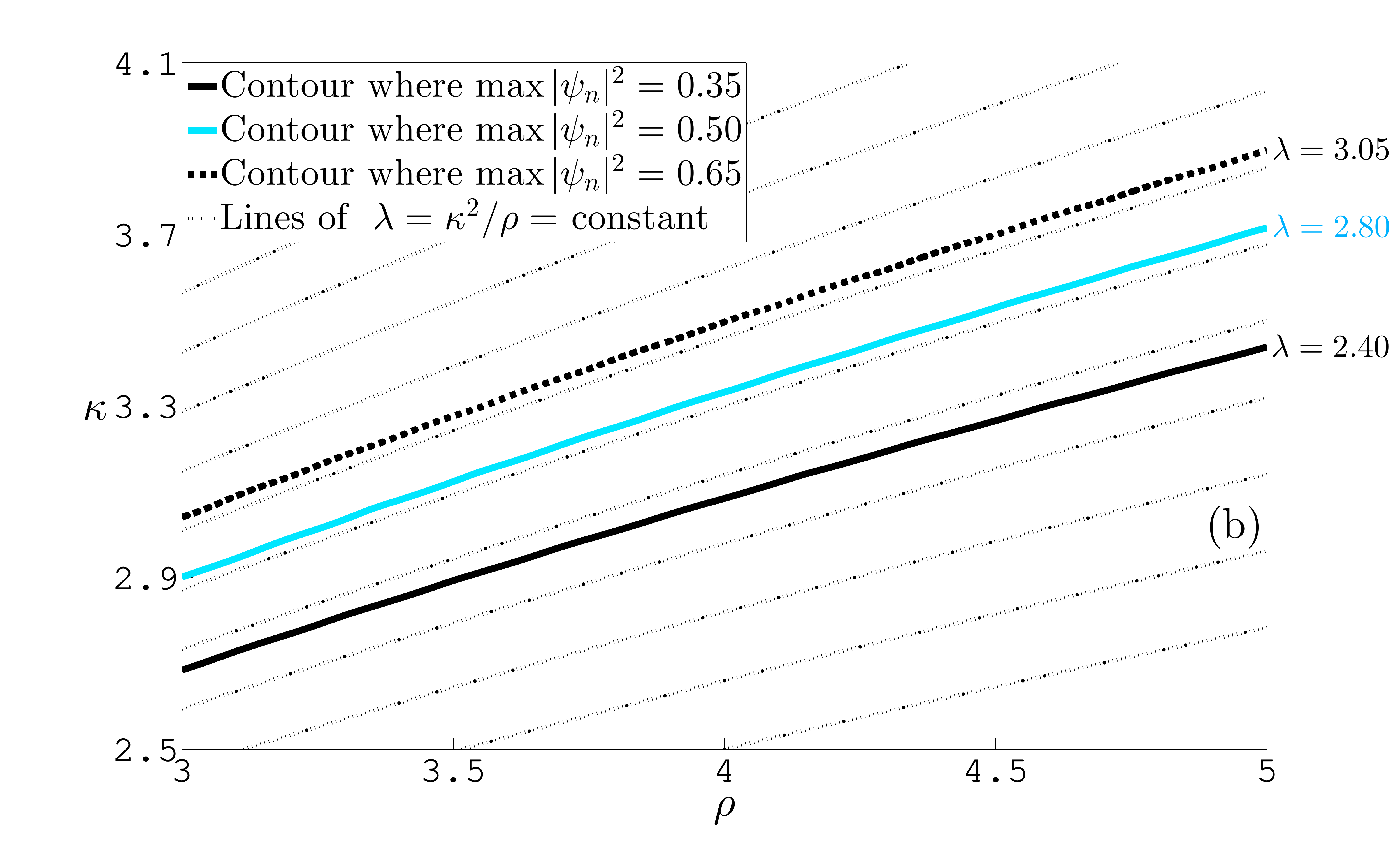} 
	\end{subfigure}%
	\caption{Dependence of $\maxpsnsq$ and $\BE^0$ on the parameter space $(\rh,\k)$. (a) $\maxpsnsq$ (positive $z$-axis) and $\BE^0$ (negative $z$-axis) as surfaces over the $(\rh,\k)$ plane. (b) Some contour lines of the $\maxpsnsq$ surface, projected onto the $(\rh,\k)$ plane. Lines of $\k^2/\rh = $ constant are included for comparison.} \label{fig_param_space}
	\end{figure}

	\subsection{Thermal equilibrium} \label{subsection_thermal_equil}

	\bg{wrapfigure}{l}{0.58\textwidth}
	\bg{center}
	\includegraphics[width=0.56\textwidth]{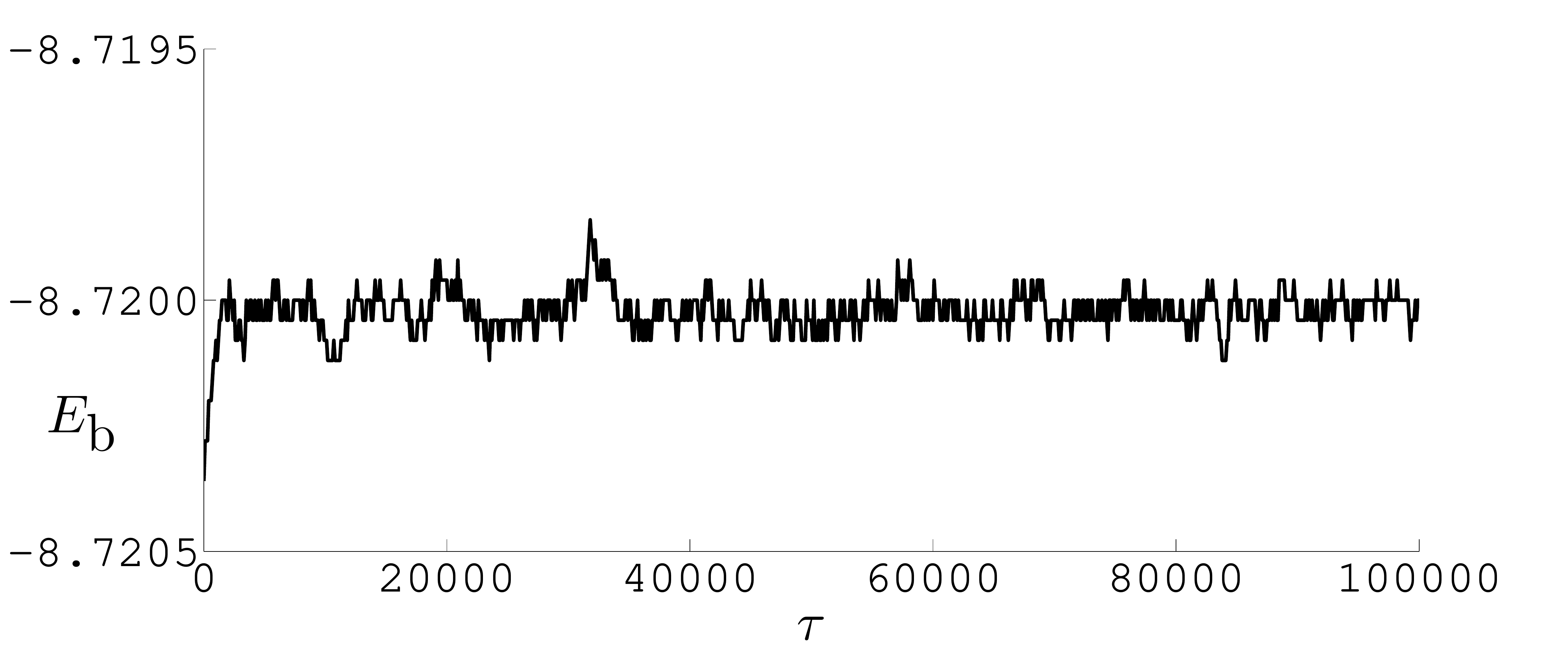}
	\end{center}
	\captionsetup{width=0.56\textwidth}
	\caption{Under non-zero stochastic forcing at $\th = 0.13$, binding energy $\BE$ of the polaron fluctuates around a stationary mean, after a rapid initial increase. $\lm = 2.80$.} \label{fig_thermalisation}
	\end{wrapfigure}
We thermalise the stationary polaron by time-evolving the system in the presence of random forces $f_n(\t)$, which take values according to \cref{thermal_term} with $\th = 0.13$ (310 Kelvin). For numerical stability, we use a small time-step of $\De \t = 0.01$. Regardless of the value of $\lm$, the polaron always settles in a \emph{quasi-stationary state}, where its binding energy, after a very small initial increase, oscillates about a steady value, and the electron probability density fluctuates around a steady configuration. This is a state of thermal equilibrium. The energy variation is always very small in magnitude, as Figure \ref{fig_thermalisation} illustrates. 
The changes in $| \ps_n |^2$ is also small, with $\maxpsnsq$ never deviating by more than $0.01$ times its initial value. These results show that our polaron is stable against thermal fluctuations acting on the lattice at physiological temperature. 


	\section{Polaron propagation induced by electric pulses} \label{section4}

With stationary polarons as initial conditions, we impose external electric fields, represented by the $\eps(\t)$ term in \cref{dimlesseqns}, and integrate our system forward in time using the RK4 method. The time-step remains $\Delta \t = 0.01$. We also set in \cref{dimlessphn} $n_0 = \bar{n}$ where the stationary $|\ps_n|^2$ profile is maximum. Neglecting $f_n(\t)$, we look for suitable $\eps$ which can facilitate polaron propagation; then in \Cref{subsection_thermal} we investigate thermal effects by turning on $f_n(\t)$. We fix $\lm =\k^2 / \rh = 2.80$, meaning that when we alter $\rh$ we also change $\k$ accordingly. This is because as we saw in \Cref{section3} the electron probability distribution of stationary polarons is parametrised only by $\lm$. We would like to study the motion of polarons with a moderate maximum localisation probability, and indeed for $\lm = 2.80$ we have $\maxpsnsq = 0.5$ in the stationary state. We discover that, in addition to $\lm$, $\kappa$ (or equivalently $\rho$) is also important to the dynamics of a non-stationary polaron, in that $(\lm = 2.80, \kappa = 3.35)$ and $(\lm = 2.80, \kappa = 3.00)$ produce very different results. We explain this further in \Cref{subsection_singlepulse,subsection_reppulse_0,subsection_thermal}.

Our numerical solutions show that several natural choices of $\eps(\t)$ produce negative results. A sinusoidal electric field causes the polaron simply to oscillate, regardless of the field's amplitude and period. Using constant $\eps$, we find that for every initial condition there exists a threshold amplitude $\eps_0$ such that no polaron displacement occurs if $\eps < \eps_0$, and if $\eps \ge \eps_0$ then the electron delocalises within several hundred units of time. Delocalisation is the phenomenon where the electron ``escapes'' the local potential well, thus destroying the polaron as its two constituent parts become decoupled. This can occur when excessive energy is imparted to the electron. Although a theoretical delocalised state is represented by a probability density with  $| \ps_n |^2 \sim \mcal{O}(1/N)$, for practical purposes we consider delocalisation to have occurred whenever the weaker condition $\maxpsnsq < 0.1$ is satisfied, at which point secondary peaks in $| \ps_n |^2$ have the same order of magnitude as $\maxpsnsq$, and this is always accompanied by a significant decay in the polaron's binding energy.
Our understanding is that the constant electric field is prone to destroying the polaron because it raises the electron energy in a sudden and continual manner. In an attempt to counter these issues, we have used a period of linear increase in $\eps(\t)$ which brings it slowly to a constant value over $\mcal{O}(10^6)$ units of time. However, the result remains that as soon as $\eps$ reaches the threshold value then the electron delocalises. This calls for a pulse-like electric potential, which peaks at a certain amplitude before resetting to zero, in theory allowing the polaron to regain stability once the peak has passed.


	\subsection{Excitation by electric pulse} \label{subsection_singlepulse}

Consider an electric pulse of the form
	\bal
	\eps(\t) = \lt\{ \begin{array}{l} A \sin^2(\pi \t / \DT), \\ 0, \end{array} \right. \quad \tn{if} \quad \left. \begin{array}{l} \t < \DT, \\ \t \geq \DT, \end{array} \right. \label{eqn_singlepulse}
	\end{align} 
where $\DT$ is the timespan of the pulse and $A$ is the amplitude. For every $\DT$ we find that there is some critical pulse amplitude $A_\tn{c}$ with the following property. 
	\bg{figure}[h!]
	\centering
	\bg{subfigure}[h]{0.45\textwidth}
	\includegraphics[width=\textwidth]{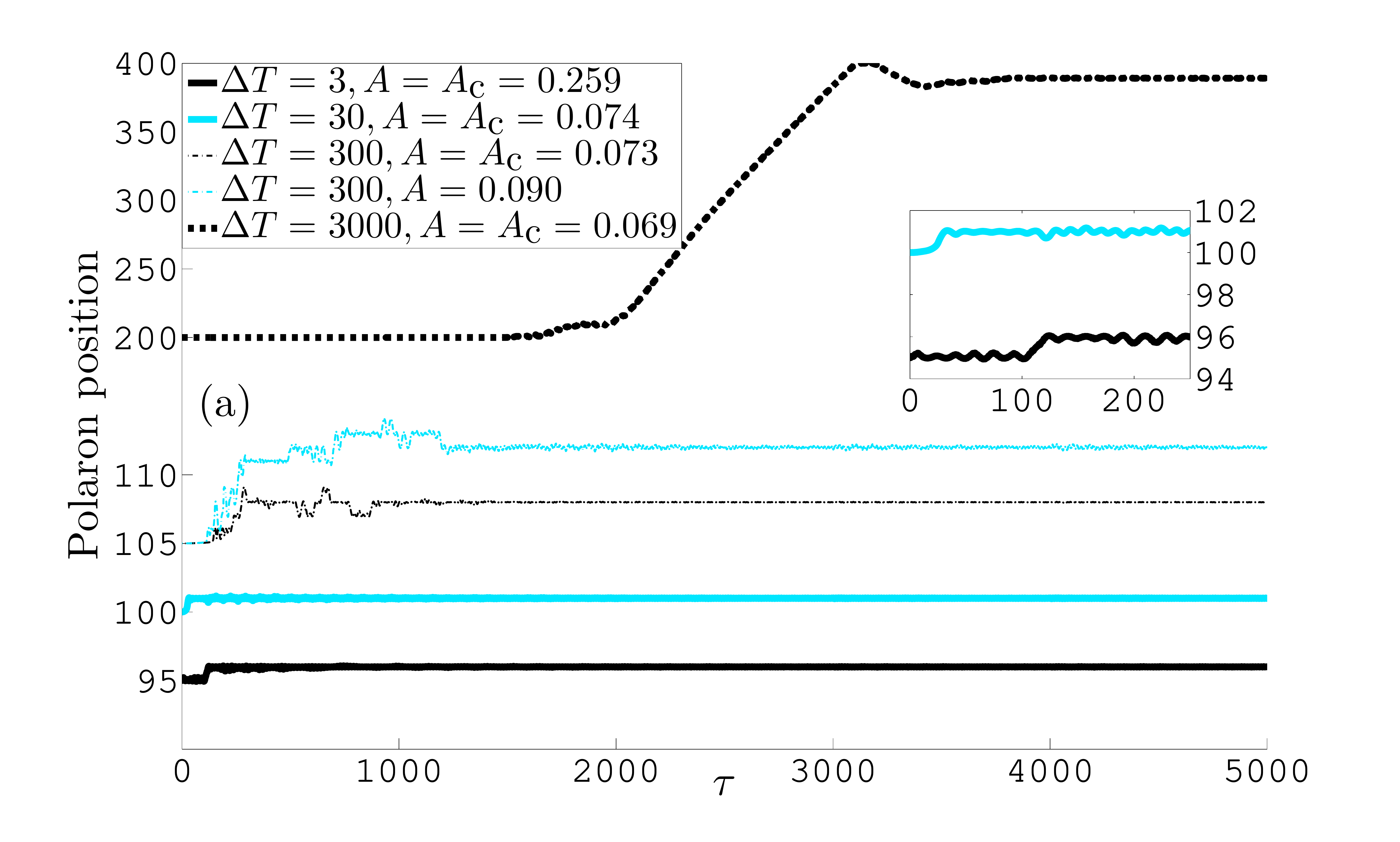}
	\end{subfigure}%
	\bg{subfigure}[h]{0.45\textwidth}
	\includegraphics[width=\textwidth]{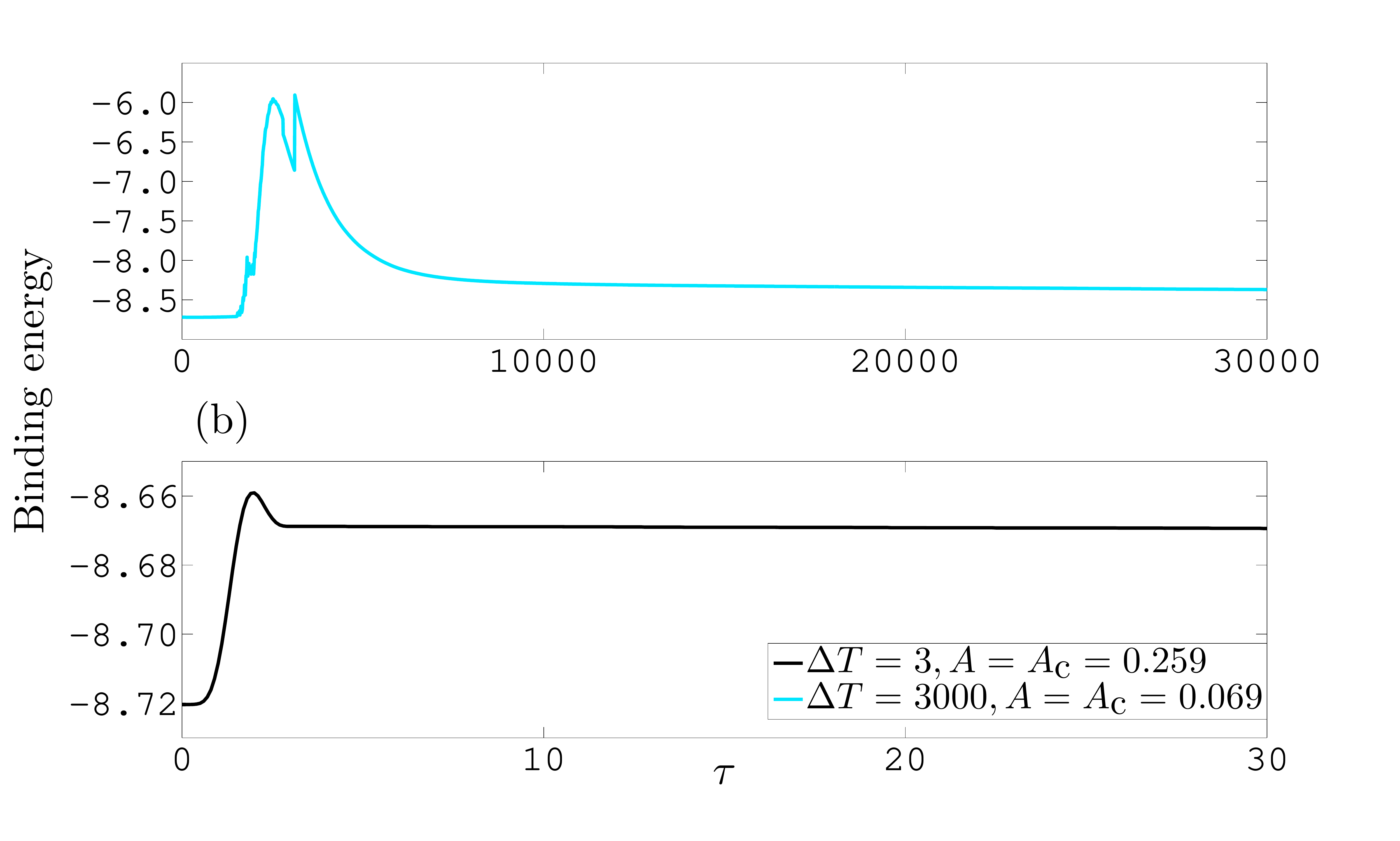} 
	\end{subfigure}%
	\captionsetup{width=0.9\textwidth}
	\caption{Polaron motion under the electric pulse with timespan $\DT$ and amplitude $A$. $\lm = 2.80$, $\kappa = 3.35$. (a) Some polaron trajectories. (b) Evolution of polaron binding energy.} \label{fig_sing}
	\end{figure}
If $A < \Ac$, the pulse causes the polaron to move away and then back to the vicinity of the initial position, before settling in a quasi-steady state of small oscillations about the initial position. The energy of the polaron is raised slightly by the pulse. If $A \geq \Ac$, the polaron moves away during the pulse but does not return, and instead settles in an oscillatory quasi-steady state some lattice sites away from its starting position. Some examples of such trajectories are shown in Figure \ref{fig_sing}(a). $A_\tn{c}$ is negatively correlated with $\DT$, which is to be expected as a longer pulse need not have as high an amplitude as a shorter one in order to impart the same amount of energy to the electron. 
	\bg{wrapfigure}{l}{0.55\textwidth}
	\bg{center}
	\includegraphics[width=0.5\textwidth]{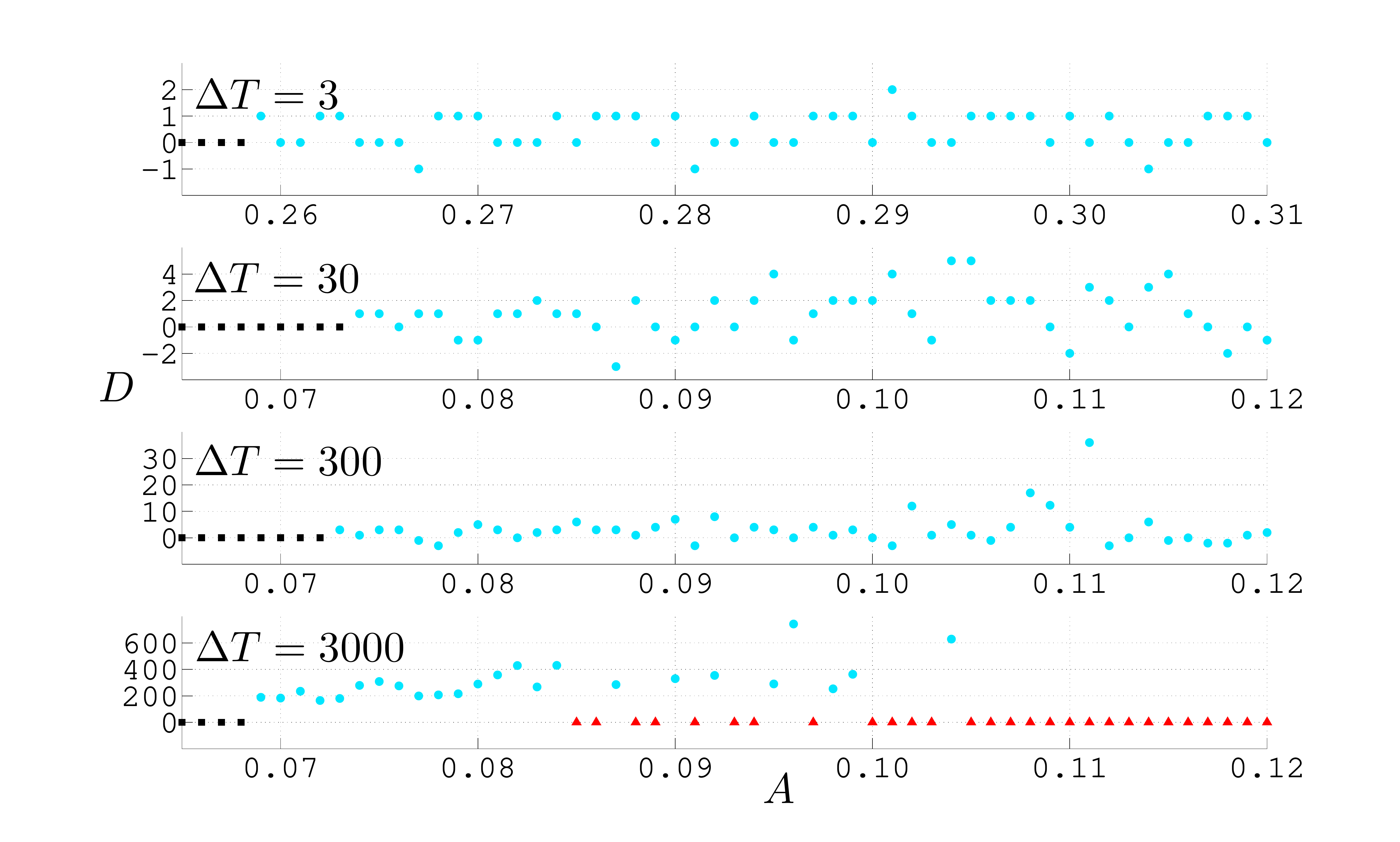}
	\end{center}
	\captionsetup{width=0.5\textwidth}
	\caption{Polaron displacement, $D$, as function of $\DT$ and $A$ [cf. \cref{eqn_singlepulse}]. $\lm = 2.80$, $\kappa = 3.35$. Squares (black) indicate zero displacement due to $A$ being too small. Triangles (red) indicate zero displacement due to delocalisation before end of pulse.} \label{fig_singlepulse1}
	\end{wrapfigure}
When there is displacement, the energy of the polaron is significantly higher in the new quasi-steady state than in the initial steady state, as Figure \ref{fig_sing}(b) illustrates. This means that the electron has picked up a large amount of energy from the pulse. Figure \ref{fig_singlepulse1} captures the way in which polaron displacement, $D$, varies as we increase $A$ beyond the critical value. We see that the relationship between $D$ and $A > \Ac$ is fairly erratic, with no clear indication of positive correlation. However, we can discern the qualitative characteristic that, the longer the timespan $\DT$, the larger the polaron displacement. This is because the system is subject to the extra energy from the pulse for longer. We note two more features of the polaron dynamics. Firstly, some combinations of $\DT$ and $A$ cause displacement of the polaron in the direction of higher electric potential, i.e. smaller $n$. This occurs when, for instance, $\DT = 30$ and $A = 0.080$ or $0.087$, even though for $0.080 < A < 0.087$ (at our resolution of $\De A = 0.001$) the polaron displacement is non-negative. Secondly, some values of $A$ cause the electron to delocalise before the end of the pulse, due to the excessive energy input. For example, when $\DT = 3000$, if $A \ge 0.105$ then delocalisation always occurs before $\t = 3000$, and some smaller values of $A$ such as 0.085 produces the same effect. 
	

	\subsection{Periodic excitations} \label{subsection_reppulse_0}

If we simply repeat the propagation-inducing single pulse over time, so that $\eps(\t)$ is periodic and takes the form $\eps(\t) = A \sin^2(\pi \t / \DT)$ for $\t \ge 0$, we obtain polaron motion which is unsustainable, in the sense that delocalisation occurs within two or three periods. Whereas for the single pulse the polaron is permanent, i.e. it remains quasi-stable once the electric field is reset to zero (as long as delocalisation has not occurred during the pulse), under the repeated pulse the polaron is transient, as it has a finite lifetime $\t_0$ at which the electron delocalises. Even though the polaron can move by several hundred lattice sites during its lifetime, its binding energy increases so rapidly that this type of polaron propagation cannot be considered an efficient transport mechanism. We understand the cause of the polaron's short lifetime to be as follows. We saw in \Cref{subsection_singlepulse} that as a pulse hits the polaron, it raises the polaron's energy, making the electron less bound to the lattice. Repeated applications of the same pulse therefore eventually decouples the electron from the lattice. Crucially, the time it takes the polaron's binding energy to re-settle at a quasi-steady value can be significantly longer than the timespan of the pulse [cf. Figure \ref{fig_sing}(b)]. This means that right at the end of the first pulse the binding energy is much higher than it would be in its quasi-steady state, and hitting the system with a second pulse straight away would raise the energy even higher. Thus, to prolong the polaron lifetime, we set $\eps(\t)$ to zero after each pulse for an amount of time equal to some $S\DT$, which we call the relaxation period, allowing the system the necessary time to settle in a new quasi-steady state before another pulse hits. We write this periodic forcing as
	\bal
	\eps(\t) = \lt\{ \begin{array}{l} A \sin^2(\pi \t / \DT), \\ 0, \end{array} \right. \quad \tn{if} \quad \left. \begin{array}{l} \t - c \lt[\lt(1+S\rt) \DT\rt] < \DT, \\ \t - c \lt[\lt(1+S\rt) \DT\rt] \geq \DT, \end{array} \right. \label{eqn_reppulse}
	\end{align} 
where $c$ is the largest integer such that $\t - c \lt[\lt(1+S\rt) \DT\rt] \geq 0$. We find that, for $\DT = 3000$, regardless of $S$, delocalisation always occurs before the end of the second pulse. For this reason, we restrict ourselves to $\DT = 3, 30$ and 300, for which $S = 10$ gives a long enough relaxation period for our purposes. 
	\bg{figure}[h!]
	\centering
	\bg{subfigure}[h]{0.5\textwidth}
	\includegraphics[width=\textwidth]{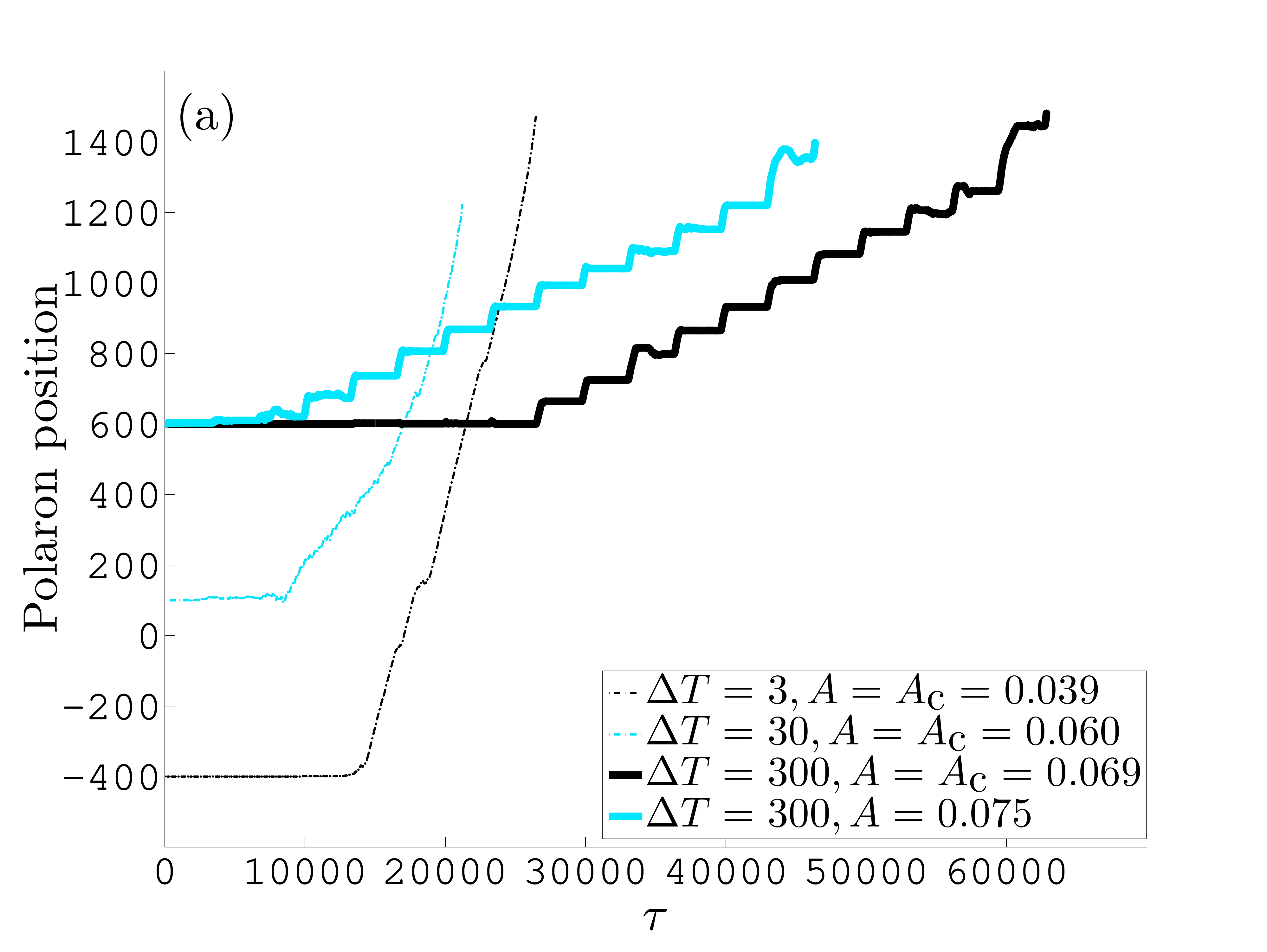}
	\end{subfigure}%
	\bg{subfigure}[h]{0.5\textwidth}
	\includegraphics[width=\textwidth]{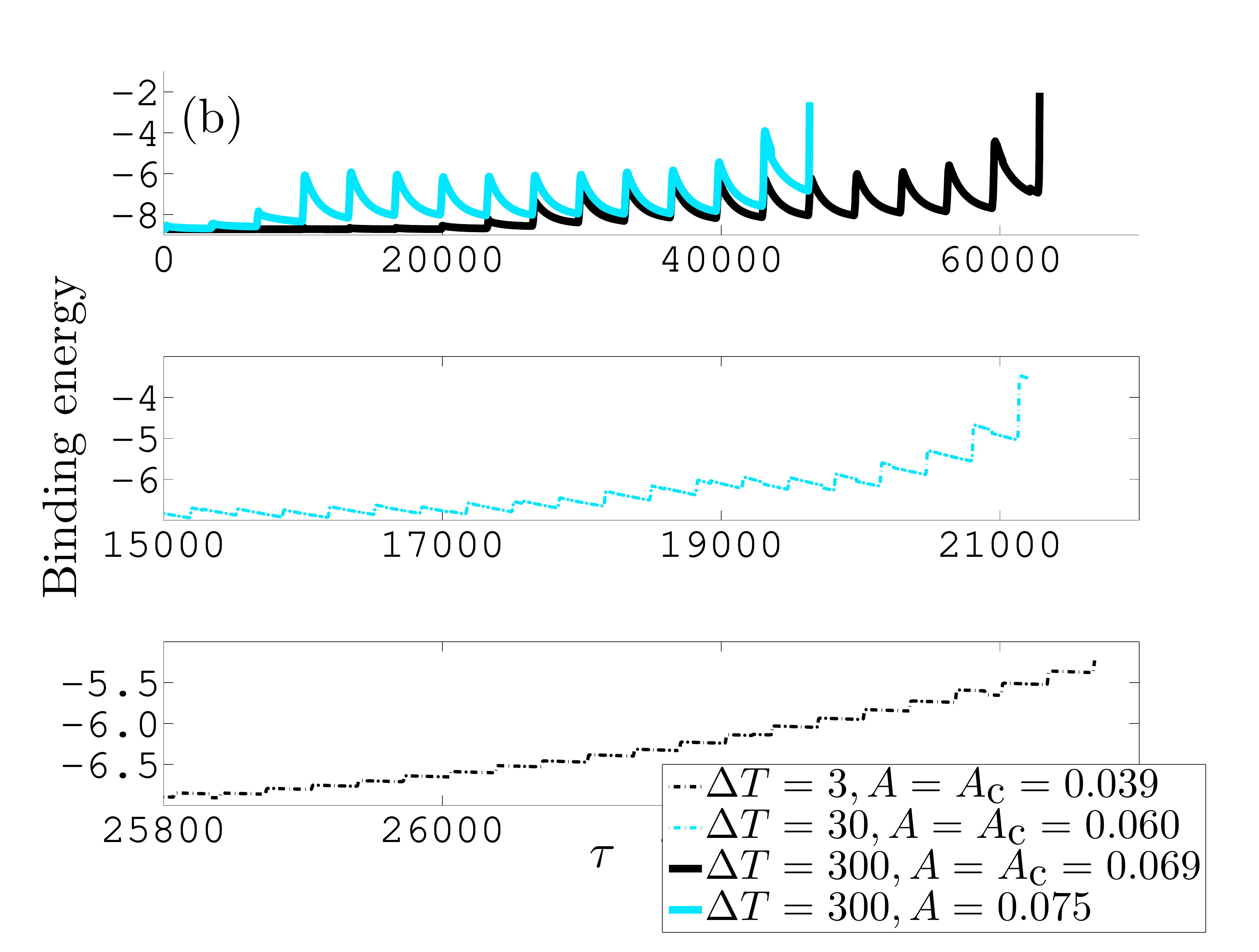} 
	\end{subfigure}%
	\captionsetup{width=0.9\textwidth}
	\caption{Polaron motion under periodic pulses with timespan $\DT$ and amplitude $A$, and relaxation period $10\DT$. $\lm = 2.80$, $\kappa = 3.35$. (a) Some polaron trajectories. (b) Evolution of polaron binding energy.} \label{fig_reppulse_long}
	\end{figure}
Figure \ref{fig_reppulse_long}(a) contains examples of polaron trajectories under the periodic forcing, and Figure \ref{fig_reppulse_long}(b) shows the corresponding evolutions of the polaron's binding energy. We see that by adding a relaxation period after each pulse we allow time for the polaron to settle into a quasi-stationary state, hence the periodic lowering of binding energy. As the polaron stabilises, its movement stalls, hence the ladder-like trajectories featuring jumps of tens of lattice sites followed by plateaus. Compared to the single pulse of \Cref{subsection_singlepulse}, periodic pulses cause much larger polaron displacements. Moreover, we saw in \Cref{subsection_singlepulse} that a pulse always raises the polaron energy (even if it does not cause a displacement), making the polaron more susceptible to moving under further pulses, which is why the critical pulse amplitude $\Ac$ for periodic pulses is lower than the $\Ac$ we saw for the single pulse. It is also why, as we see in Figure \ref{fig_reppulse_long}(a), at $A = A_\tn{c}$ the polaron does not begin to move until several pulses have hit. 

Figure \ref{fig_traj3d} is a visualisation of the way a polaron moves during a pulse and settles afterwards. 
	\begin{figure}
	\centering
	\begin{minipage}{.4\textwidth}
	\centering
	\includegraphics[width=\linewidth]{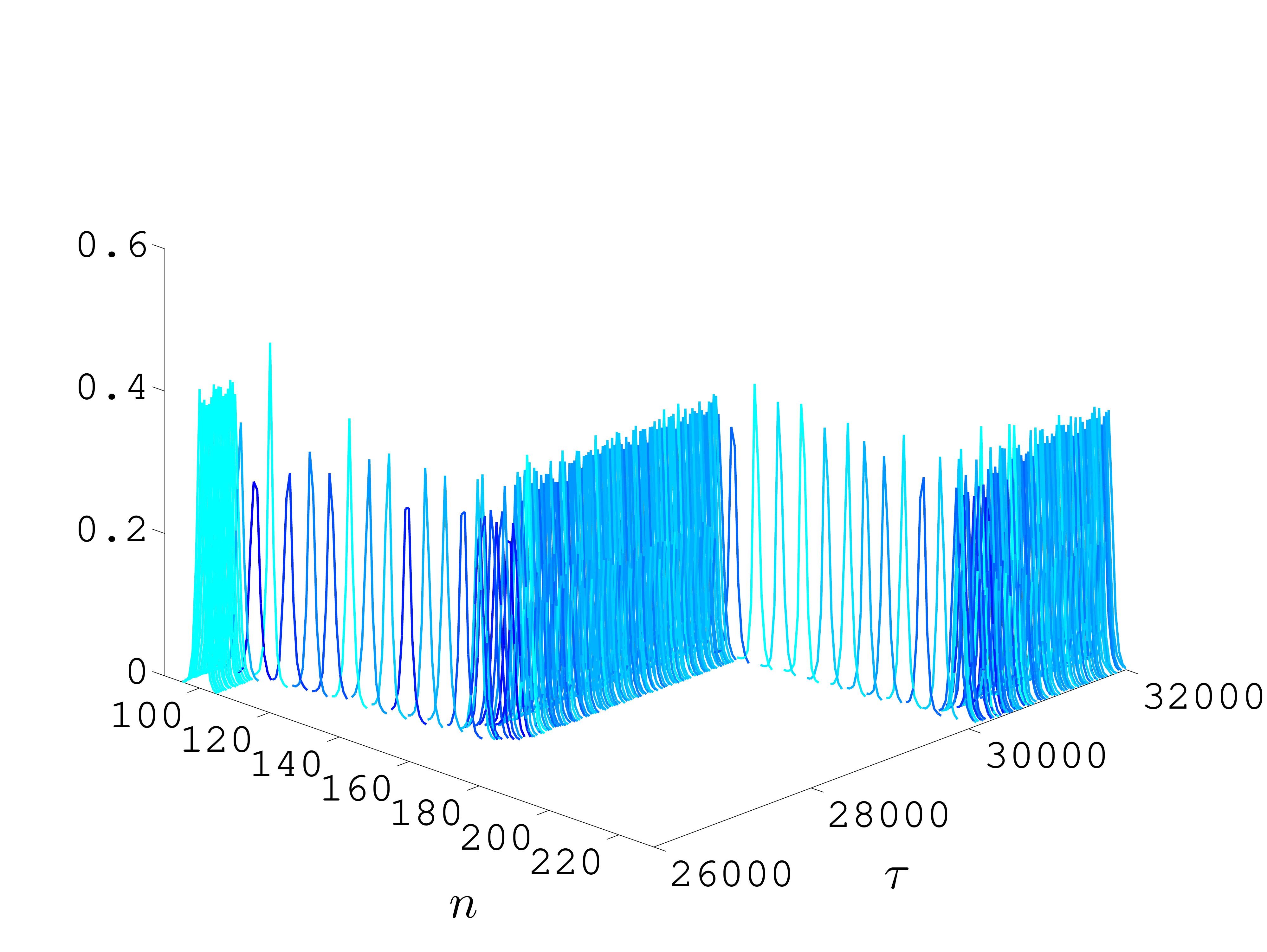}
	\captionsetup{width=0.9\linewidth}
	\caption{Evolution of the electron probability density, $| \ps_n |^2$, under periodic pulses with $\DT = 300, A = A_\tn{c} = 0.069$. $\lm = 2.80, \k = 3.35$. Each curve is a $|\ps_n|^2$ profile at some $\t$. (The broader the density, the darker the shade.)} \label{fig_traj3d}
	\end{minipage}%
	\begin{minipage}{.6\textwidth}
	\centering
	\includegraphics[width=\linewidth]{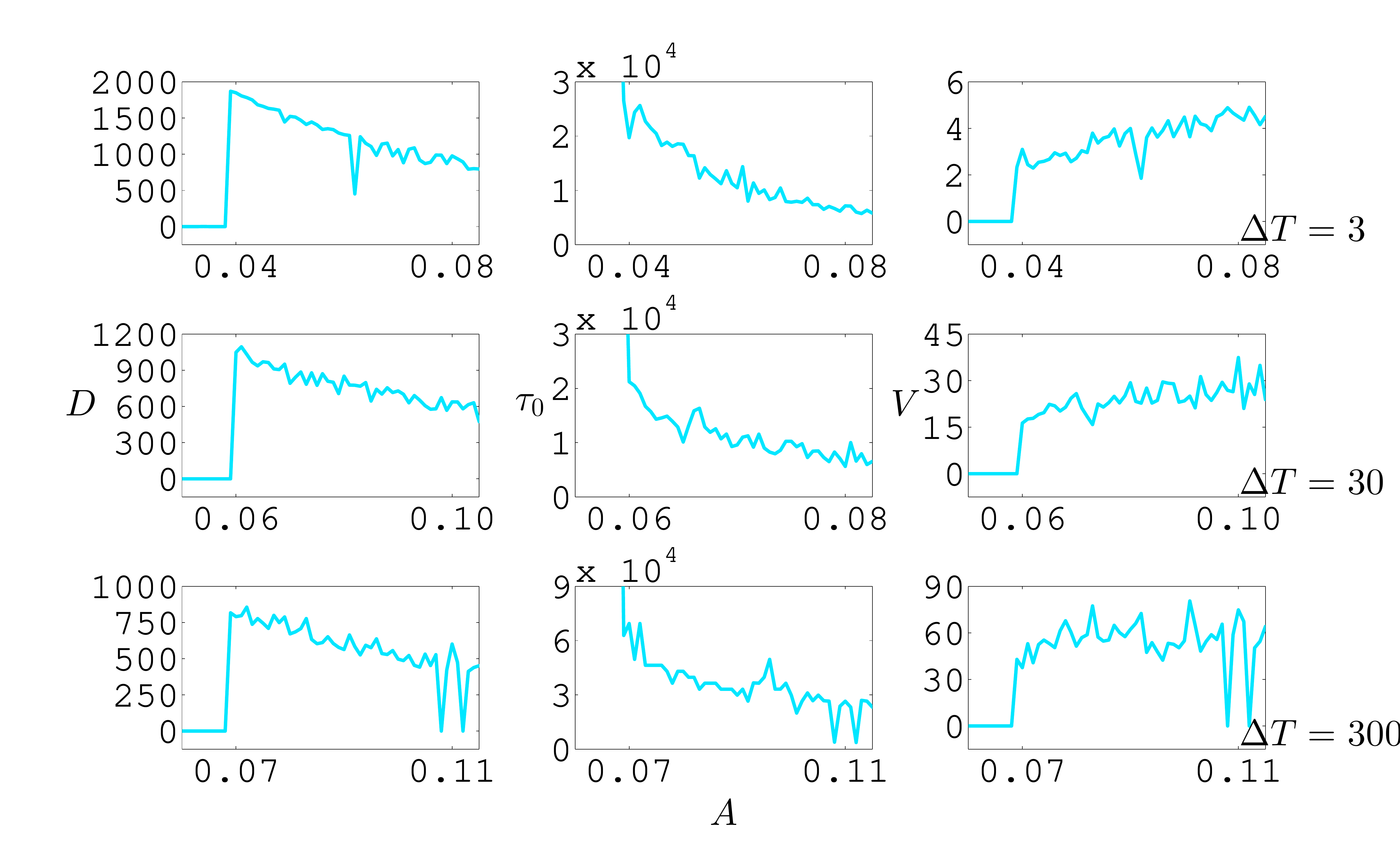}
	\captionsetup{width=0.9\linewidth}
	\caption{Polaron displacement, $D$, lifetime, $\t_0$, and displacement per pulse, $V$, as functions of $\DT$ and $A$ [cf. \cref{eqn_reppulse}]. $\lm = 2.80$, $\kappa = 3.35$, $S = 10$.} \label{fig_reppulse1}
	\end{minipage}
	\end{figure}
The polaron's size, represented by the breadth of the electron probability function, oscillates during the pulse, and after each pulse the probability density always becomes broader than it was before. Examining Figure \ref{fig_reppulse1}, we recognise a clear negative correlation between the polaron's displacement $D$ and the pulse amplitude $A$ for $A > A_\tn{c}$, as opposed to the erratic relationship between $D$ and $A$ in the case of a single pulse [cf. Figure \ref{fig_singlepulse1}]. The negative correlation can be explained as follows. 
The polaron's lifetime is negatively correlated with $A$, as a stronger pulse raises the polaron energy by a larger amount and its repeated application causes delocalisation more quickly. Meanwhile, the displacement per pulse tends to increase with $A$, as we see in the 3\tsups{rd} column of subfigures in Figure \ref{fig_reppulse1}, but this increase is small compared to the decay in polaron lifetime. As a result, total displacement over the polaron's lifetime is a decreasing function of $A$, for $A > A_\tn{c}$. This has the implication that $A_\tn{c}$ is not only the critical amplitude, but also in a sense the \emph{optimal amplitude}, and it induces the largest displacement. We note in addition that under the periodic forcing the polaron propagation is \emph{directed}, meaning all combinations of $\DT$ and $A$ cause displacements in the same direction, which was not the case under single-pulse forcing. 
	\bg{figure}[h!]
	\centering
	\bg{subfigure}[h]{0.4\textwidth}
	\includegraphics[width=\textwidth]{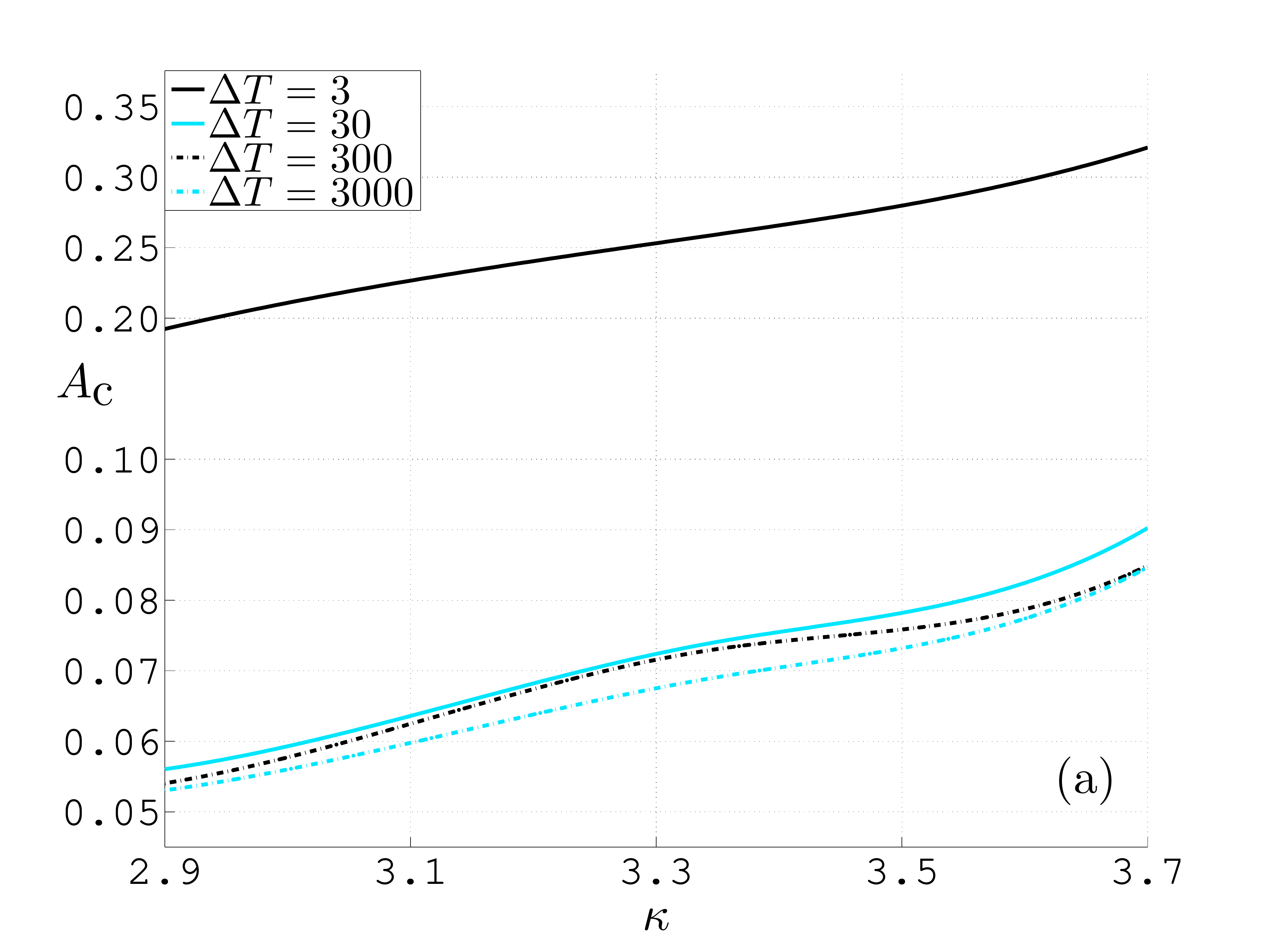}
	\end{subfigure}%
	\bg{subfigure}[h]{0.4\textwidth}
	\includegraphics[width=\textwidth]{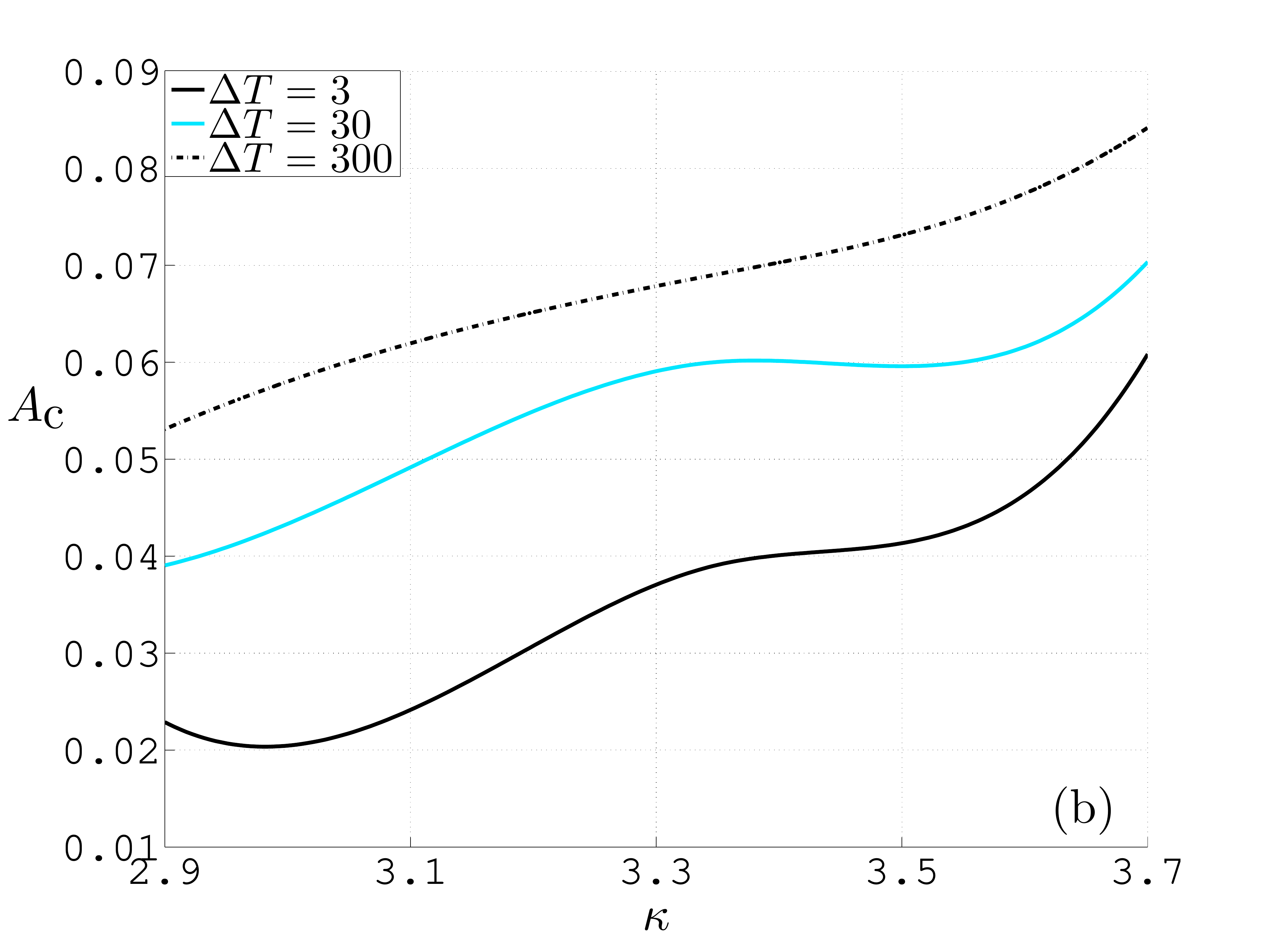} 
	\end{subfigure}%
	\captionsetup{width=0.8\textwidth}
	\caption{$A_\tn{c}$ as a function of $\kappa$, parametrised by $\DT$. $\lm = 2.80$. (a) Single-pulse forcing by \cref{eqn_singlepulse}. (b) Periodic forcing by \cref{eqn_reppulse}.} \label{fig_kappa}
	\end{figure}
Our results show that using different combinations of $\rh$ and $\k$, keeping $\lm = \k^2 / \rh$ fixed, produces figures which are characteristically similar to Figure \ref{fig_reppulse1}, showing polaron displacement and lifetime as decreasing functions of $A$ for $A > A_\tn{c}$. However, the value of $A_\tn{c}$ is dependent on more than just $\lm$. Figure \ref{fig_kappa} exhibits the dependence of $A_\tn{c}$ on the electron-lattice coupling strength $\kappa$, with $\lm = 2.80$ fixed. It shows that the more strongly coupled our system is, the more difficult it becomes to displace a polaron using pulse-like electric fields, in the sense that a larger amplitude is required to cause the onset of polaron propagation. Comparing Figure \ref{fig_kappa}(a) and Figure \ref{fig_kappa}(b), we see that periodic forcing requires a much lower pulse amplitude to achieve polaron displacement than single-pulse forcing, particularly when $\DT$ is small. Indeed, when $\DT = 3$, $A_\tn{c}$ for the periodic forcing is an order of magnitude smaller than that for the single pulse.


	\subsection{Thermal stability} \label{subsection_thermal}
	
We study the effect of stochastic forcing from the environment by first evolving the system of \cref{dimlesseqns} under a non-zero $f_n(\t)$ and $\eps(\t) = 0$ until it reaches thermal equilibrium as we described in \Cref{subsection_thermal_equil}, and then turning on $\eps(\t)$ as per \Cref{subsection_singlepulse,subsection_reppulse_0}. 
	\begin{figure}[h]
	\centering
	\begin{minipage}{.5\textwidth}
	\centering
	\includegraphics[width=\linewidth]{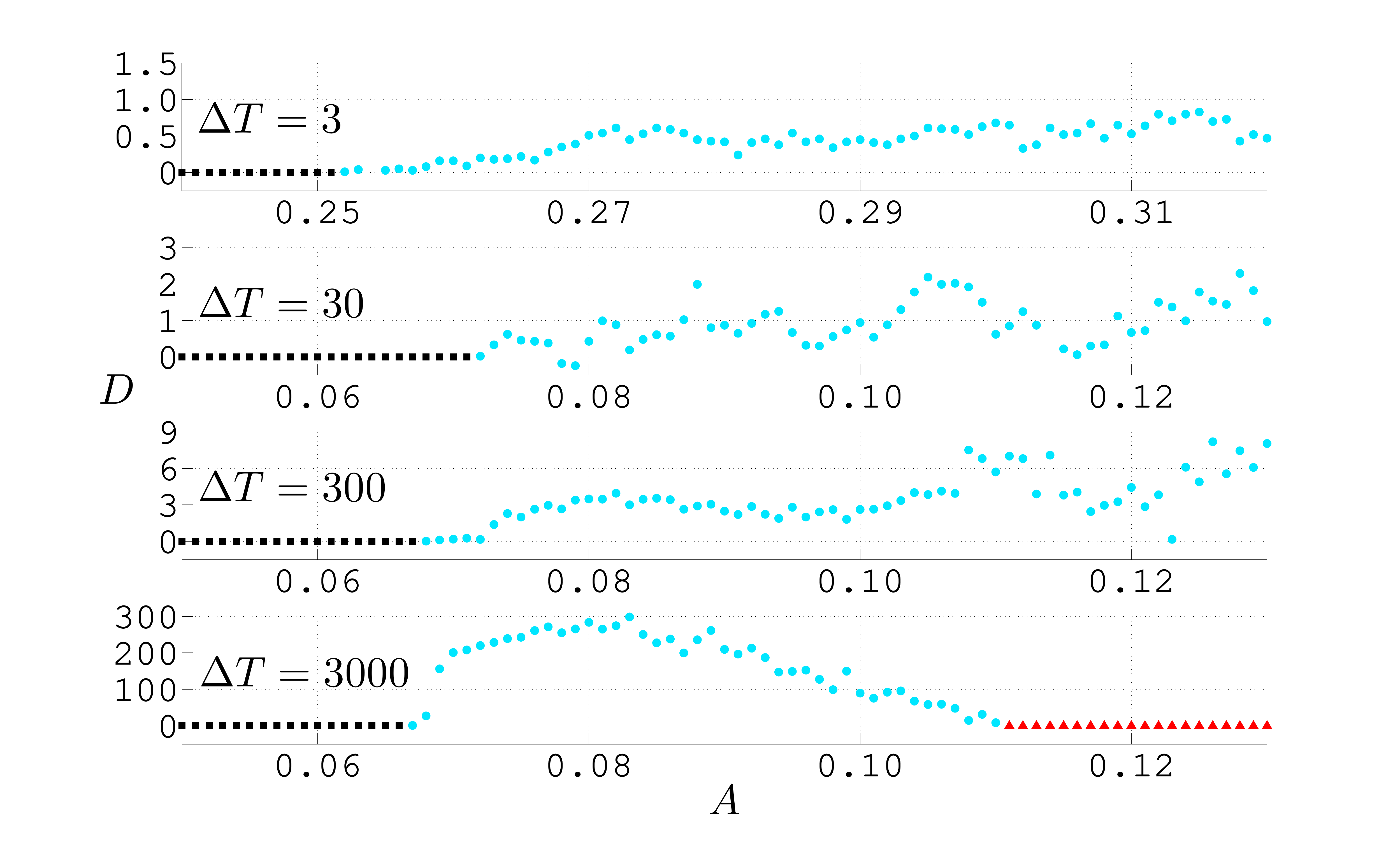}
	\captionsetup{width=0.9\linewidth}
	\caption{Mean displacement $D$ as function of $\DT$ and $A$; single pulse, $\th = 0.13$, $\lm = 2.80$, $\kappa = 3.35$. Squares (black) indicate zero displacement due to $A$ being too small. Triangles (red) indicate zero displacement due to delocalisation before end of pulse.} \label{fig_thermal1}
	\end{minipage}%
	\begin{minipage}{.5\textwidth}
	\centering
	\includegraphics[width=\linewidth]{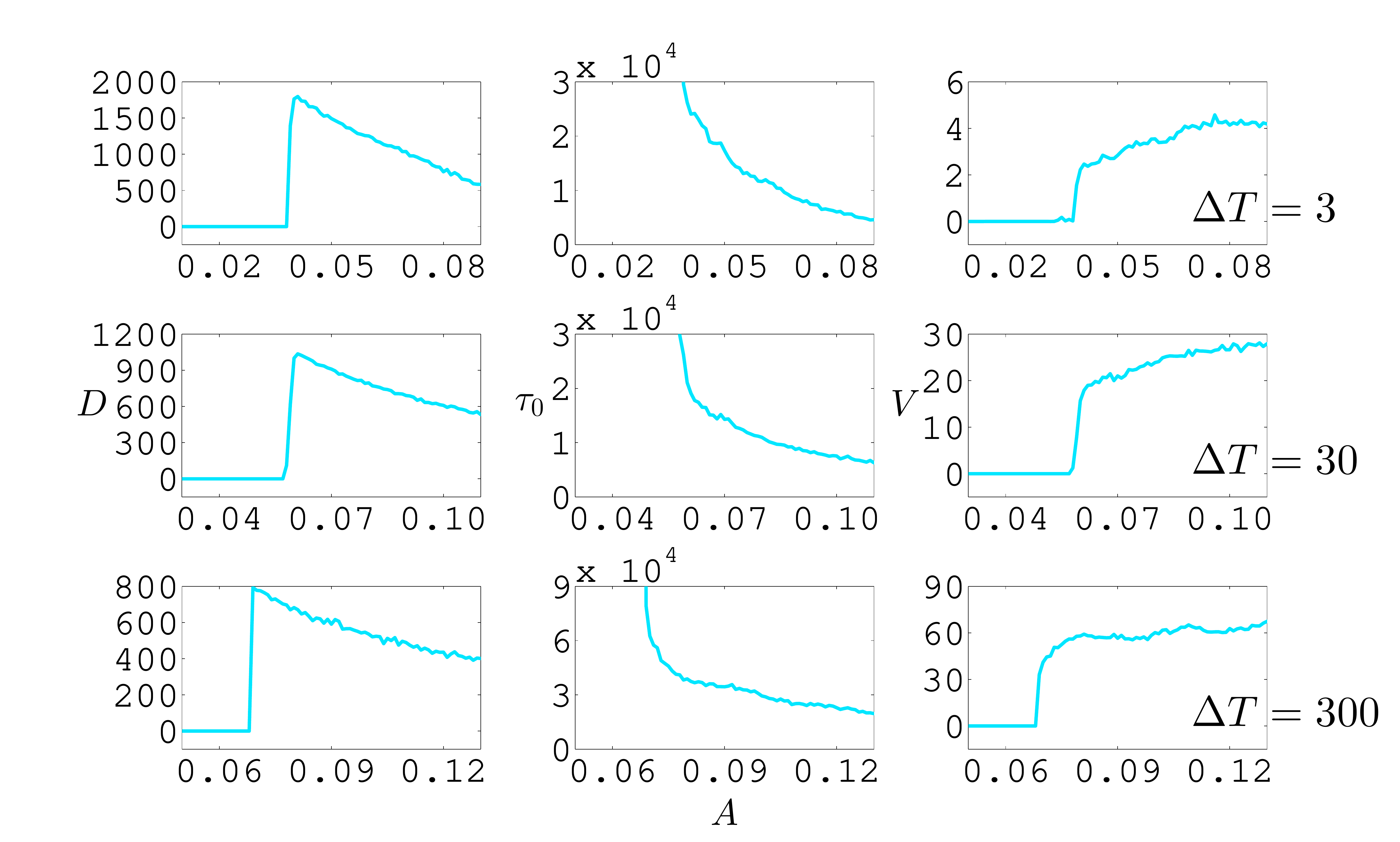}
	\captionsetup{width=0.9\linewidth}
	\caption{Mean values of displacement, $D$, lifetime, $\t_0$, and displacement per pulse, $V$, as functions of $\DT$ and $A$; periodic pulses, $\th = 0.13$, $\lm = 2.80$, $\kappa = 3.35$, $S = 10$.} \label{fig_thermal2}
	\end{minipage}
	\end{figure}	
The stochastic term exists due to thermal energy $\th := k_B T / (\hb \O)$ and satisfies \cref{thermal_term}. We fix $\th = 0.13$ which corresponds to physiological temperature (310K). For each set of parameter values $(\lm, \k, \DT, A)$, we have run 100 numerical simulations and taken the mean values of key scalar quantities associated with the polaron motion, namely its displacement and, in the case of periodic pulse-like electric fields, lifetime and displacement per pulse. Figure \ref{fig_thermal1} presents the way in which the polaron displacement $D$ depends on the amplitude $A$ of the single-pulse forcing, under random thermal fluctuations, and it may be compared directly to Figure \ref{fig_singlepulse1}, for which $f_n = 0$. 
	\bg{figure}[h!]
	\centering
	\bg{subfigure}[h]{0.4\textwidth}
	\includegraphics[width=\textwidth]{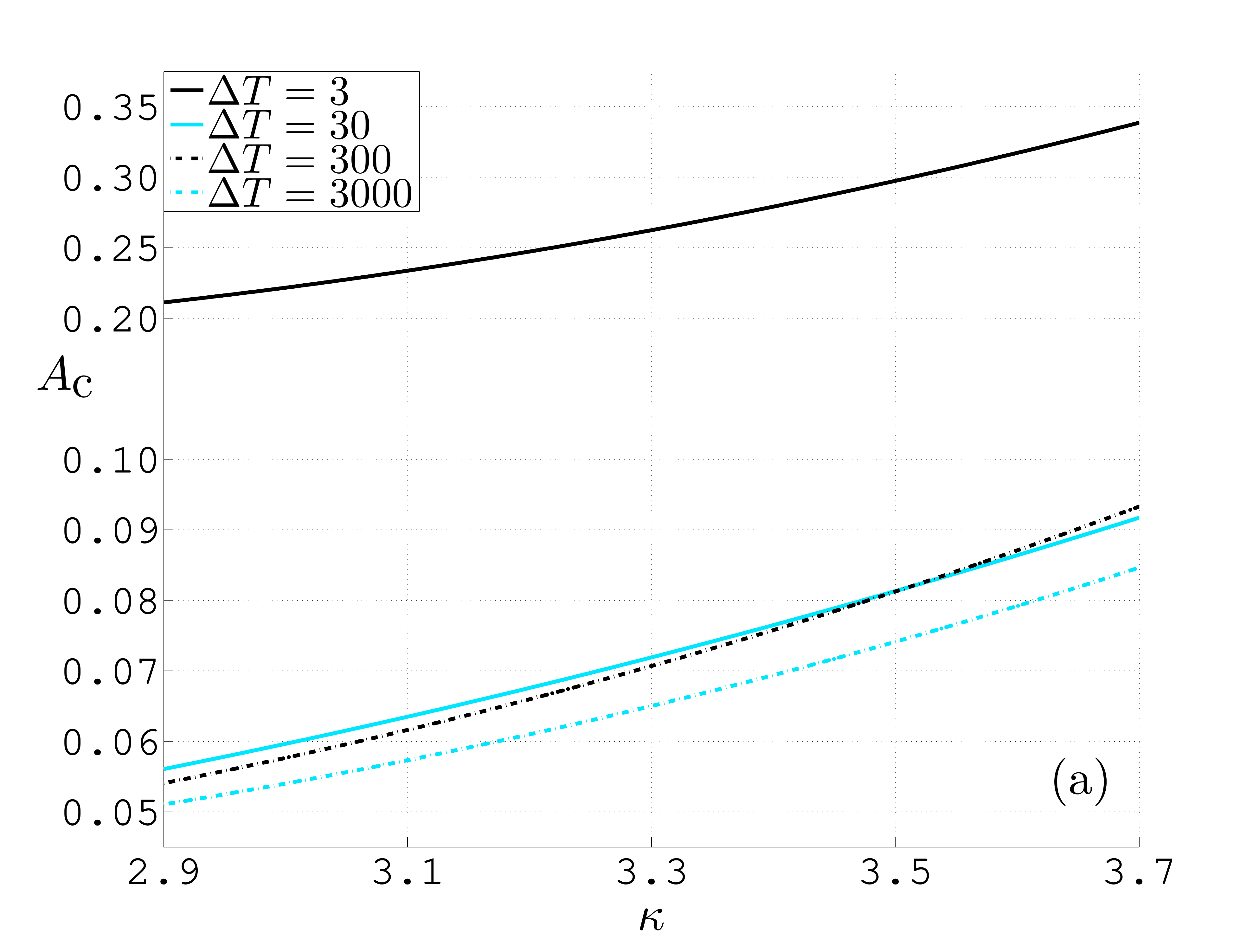}
	\end{subfigure}%
	\bg{subfigure}[h]{0.4\textwidth}
	\includegraphics[width=\textwidth]{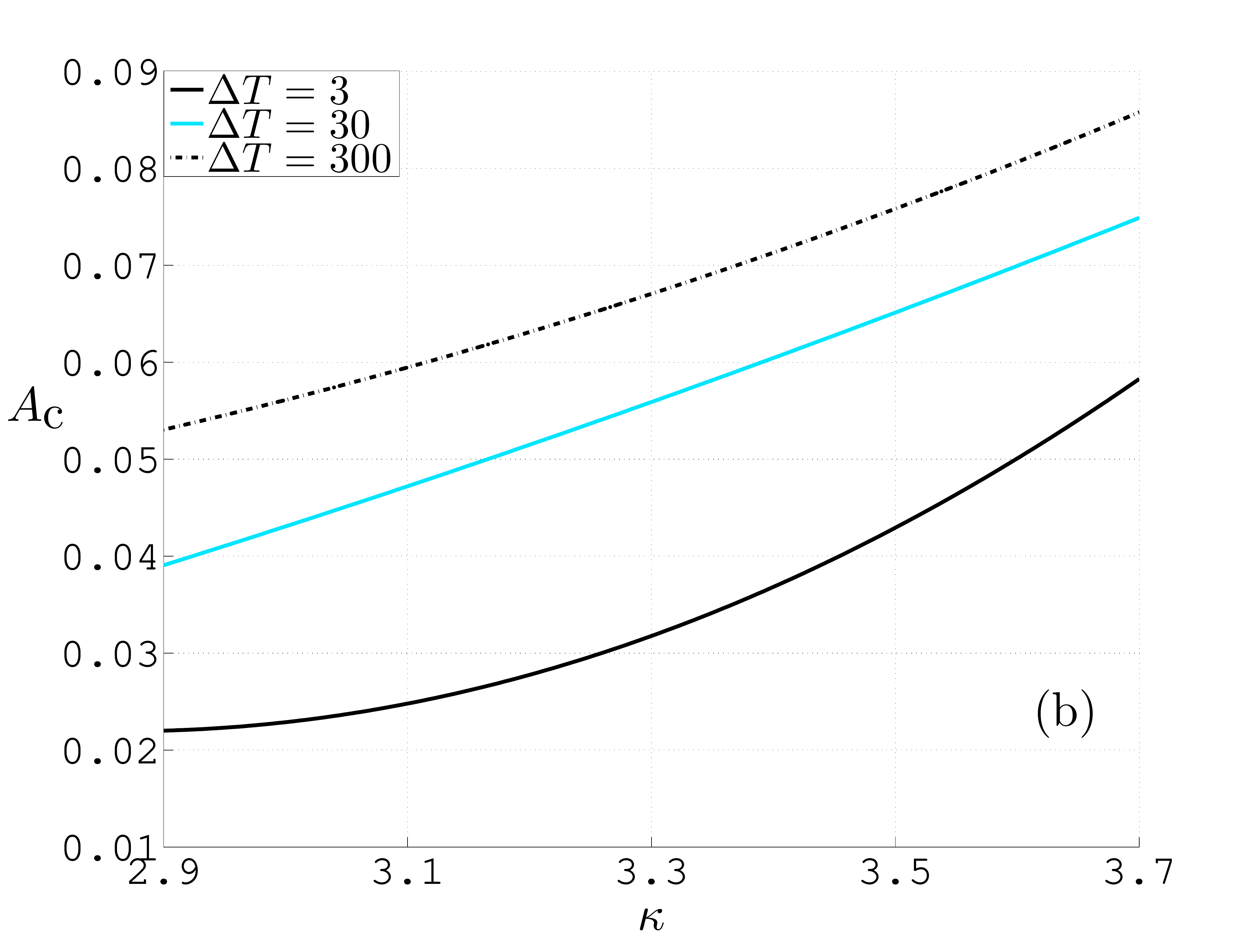} 
	\end{subfigure}%
	\captionsetup{width=0.8\textwidth}
	\caption{$A_\tn{c}$ as a function of $\kappa$, parametrised by $\DT$. $\lm = 2.80$, thermal energy $\th = 0.13$. (a) Single-pulse forcing by \cref{eqn_singlepulse}. (b) Periodic forcing by \cref{eqn_reppulse}.} \label{fig_kappa_thermal}
	\end{figure}
One of the notable effects of the stochastic forcing is making the polaron motion more directed, as Figure \ref{fig_thermal1} shows few combinations of $\DT$ and $A$ causing negative displacements. Figure \ref{fig_thermal2} shows results of combining the periodic excitation of \cref{eqn_reppulse} with stochastic forcing, and we can compare it with Figure \ref{fig_reppulse1} for which $f_n = 0$. The phenomenon that polaron displacement per pulse is an increasing function of $A$ is more clearly seen when stochastic forces are in play. The correlation between total displacement and $A$ for $A > A_\tn{c}$, and between polaron lifetime and $A$, both of which are negative, are also stronger under thermal fluctuations. Due to this negative correlation, the critical pulse amplitude $A_\tn{c}$ serves also as optimal amplitude, being the pulse strength that induces the largest displacement. Polaron propagation is strongly directed, in that the mean value over 100 numerical simulations of total displacement is several hundred lattice sites in the positive direction. We also note the stabilising effect of the thermal fluctuations, which is evidenced by the smoothness of the displacement function $D$ of $A$, as opposed to the jagged $D$-versus-$A$ curves for $f_n = 0$ [cf. Figure \ref{fig_reppulse1}] which exhibit significant dips at some values of $A$. Crucially, the polaron's lifetime is not at all reduced in the thermal environment compared to its lifetime under zero random forcing. The stochastic forcing also has a small effect on the critical pulse amplitudes, $\Ac$, namely that it slightly decreases $\Ac$, meaning it is easier to displace a polaron when thermal fluctuations take place. Comparing Figure \ref{fig_kappa_thermal} to Figure \ref{fig_kappa} reveals this difference. Figure \ref{fig_kappa_thermal} also reveals that the stochastic forcing has little bearing on the way in which $\Ac$ increases with the electron-lattice coupling strength, $\k$. Finally, we note that $\Ac$ increases with $\DT$ when the forcing is periodic, whilst the correlation is negative under the single-pulse forcing.
	

	\section{Discussions and conclusions} \label{section5}

We have put forward a model for the interaction between an electron and \amideone~vibrations in a linear polypeptide. We have shown that the interaction can result in stationary polarons whose probability density and binding energy depend on an effective coupling parameter $\lm = \k^2/\rh$, where $\rh$ and $\k$ represent respectively the adiabaticity and coupling strength in the electron-amide system. In particular, the maximum value of the probability density function $\abs{\ps_n}^2$ is proportional to $\lm$, and a more localised $\abs{\ps_n}^2$ represents a state with lower energy. To induce the propagation of a polaron with a moderate size from its stationary state, we have used an external excitation in the form of a squared-sinusoidal electric pulse, after constant and permanently sinusoidal electric fields produced negative results. The timespan of a pulse ranged between 3 and 3000 units, which correspond respectively to 0.01ps and 10ps, ensuring that the dynamical evolution of our polaron under a single pulse takes place within the picosecond timescale reported in \cite{Edler2002} as the lifetime of \amideone~excitations.  We have discovered that for every pulse timespan $\DT$ there exists a pulse amplitude $A = \Ac$ which is critical, in the sense that an excitation displaces the polaron if and only if $A \ge \Ac$. When displacement occurs, the polaron typically moves along the polypeptide by a distance which is positively correlated with $\DT$, before settling in a quasi-stationary state, with its energy raised compared to its pre-excitation level. By repeating the electric pulse periodically in time, with a sufficiently long relaxation period between pulses to allow the polaron to settle, we have found that the polaron can remain intact for up to tens of pulses, as each pulse causes a displacement in the same direction along the polypeptide. For sufficiently small $\DT$, there can be many pulses within the characteristic \amideone~lifetime, causing a total displacement by tens of peptide units. For $A > \Ac$, the total displacement and polaron lifetime are both decreasing functions of $A$, even though the displacement per pulse increases with $A$. Moreover, while fixing $\lm$ we have varied the coupling strength $\k$ in order to investigate its effect on the polaron dynamics, and we have found that $\Ac$ is positively correlated with $\k$. Our results show that polaron propagation induced by pulse-like electric potentials can occur at physiological temperatures. Indeed, thermally-induced stochastic forcing on the peptide units have a stabilising effect on the system, and it promotes directed transport towards one end of the polypeptide over the other. Over the lifetime of a polaron, it can move along the polypeptide by up to hundreds of units.

Our model puts the electron-lattice system in a coherent state characterised by electron creation and annihilation operators, as well as position and momentum operators for the lattice points. We have studied an alternative model where the lattice points are described by classical variables, and the dynamical equations for that system are the same as \cref{dimlesseqns} except that $\o(\t)$ and $\eta(\t)$ do not appear. The results we obtained from the alternative model were almost indistinguishable from that which we have presented in this paper, except for the fact that the quantum correction terms $\o(\t)$ and $\eta(\t)$ appear to have the effect of lowering the critical amplitude of electric pulses by up to 5\%, when thermal fluctuations are neglected. With thermal forcing taken into account, results under the two models are essentially identical. This agrees with findings in \cite{Cruzeiro-Hansson1995}.

In our model we have thermalised the system classically using a Langevin term for the lattice field. F\"orner \cite{Forner1993} argued that one should use a quantum thermalisation of the system and showed that when doing so the thermal stability of the Davydov soliton is smaller than what is predicted by a classical thermalisation. While it would be interesting to do a full quantum thermalisation for our model, we believe it would make very little difference as in our case the exchange energy $J_1$ is three orders of magnitude larger than in the Davydov model and approximately 30 times larger than the thermal energy at 300 Kelvin.

The strengths of electric fields involved in our model ranged from 0.5 to 20 millivolts per angstrom. In the biological cell, if a pair of opposite charges on either side of the plasma membrane spontaneously localised, the resulting dipole would create an electric field in the centre of the membrane with strength $\sim 1/(\eps_\tn{r} d^2)$, where $\eps_\tn{r}$ and $d$ are respectively the relative permittivity and width of the membrane \cite{Jackson1999,Li2013}. Taking $\eps_\tn{r} = 5$ \cite{Weaver2003} and $d = 80$\ang \cite{McCaughan1980,Hochmuth1983}, this electric field has strength 3.6 millivolts per angstrom, which is within the range of values we have used in our model. Moreover, observations of hundred-femtosecond charge separation in biological complexes have been reported \cite{Gauduel1987,Zhong2001}, the timescale of which matches the timespan of electric pulses we have considered. It is therefore conceivable that the pulse-like electric potentials in our model could naturally occur and induce directed electron transport along polypeptides across the cell membrane.



\end{document}